# *Thermodynamic Paradox and Non-Hermitian Topological Singularities*

*Mário G. Silveirinha*[(1)*]

[(1)] *University of Lisbon–Instituto Superior Técnico and Instituto de Telecomunicações, Avenida Rovisco Pais, 1, 1049-001 Lisboa, Portugal*

**Abstract**

Unidirectional modes in magnetically biased electromagnetic systems have long been associated with a thermodynamic paradox: the absence of counter-propagating channels may produce field "hotspots" that can act as unphysical sinks of thermal radiation. Here we revisit this problem and show that, surprisingly, material dissipation alone cannot fully regularize the singular behavior of the normal modes of a nonreciprocal cavity. We demonstrate that the paradox is resolved by nonlocal effects, which suppress the material response at short wavelengths and eliminate field singularities altogether. Our analysis reveals a fundamental link between nonlocality, topology, and thermodynamic consistency, showing that real-space singularities and ill-defined topologies go hand in hand, even in strongly dissipative platforms. These findings clarify the physical origin of the paradox and establish nonlocality as a natural and robust mechanism for its resolution, opening new avenues to explore the intertwined roles of topology and nonlocality in passive nonreciprocal photonics.

* To whom correspondence should be addressed: E-mail: mario.silveirinha@tecnico.ulisboa.pt

# I. The Paradox

Unidirectional propagation in nonreciprocal photonic systems has long been recognized as one of the most striking consequences of magneto-optical biasing. First observed in ferrite waveguides in the 1950s [1-4] and later rediscovered in the context of topological photonics [5-8], such modes form the basis of a broad range of devices such as electromagnetic isolators and circulators. Yet their very existence also poses a paradox: in closed geometries, the absence of a counter-propagating channel may prevent thermal energy from circulating consistently, seemingly conflicting with the requirements of thermodynamic equilibrium [9-11]. This challenges the fluctuation–dissipation framework that underpins noise and energy transport in nonreciprocal systems.

The purpose of this work is to revisit this paradox through a systematic modal analysis and to identify the precise physical mechanism that eliminates the associated real-space field singularities. For clarity, we focus on the simplest configuration where the issue arises: a closed metallic cavity partially filled with a magnetically biased ferrite slab, sketched in Fig. 1a. The cavity can be viewed as a waveguide supporting propagation along the $+x$ direction and terminated by metallic walls. Nonetheless, the analysis and conclusions are general and extend readily to other nonreciprocal platforms.

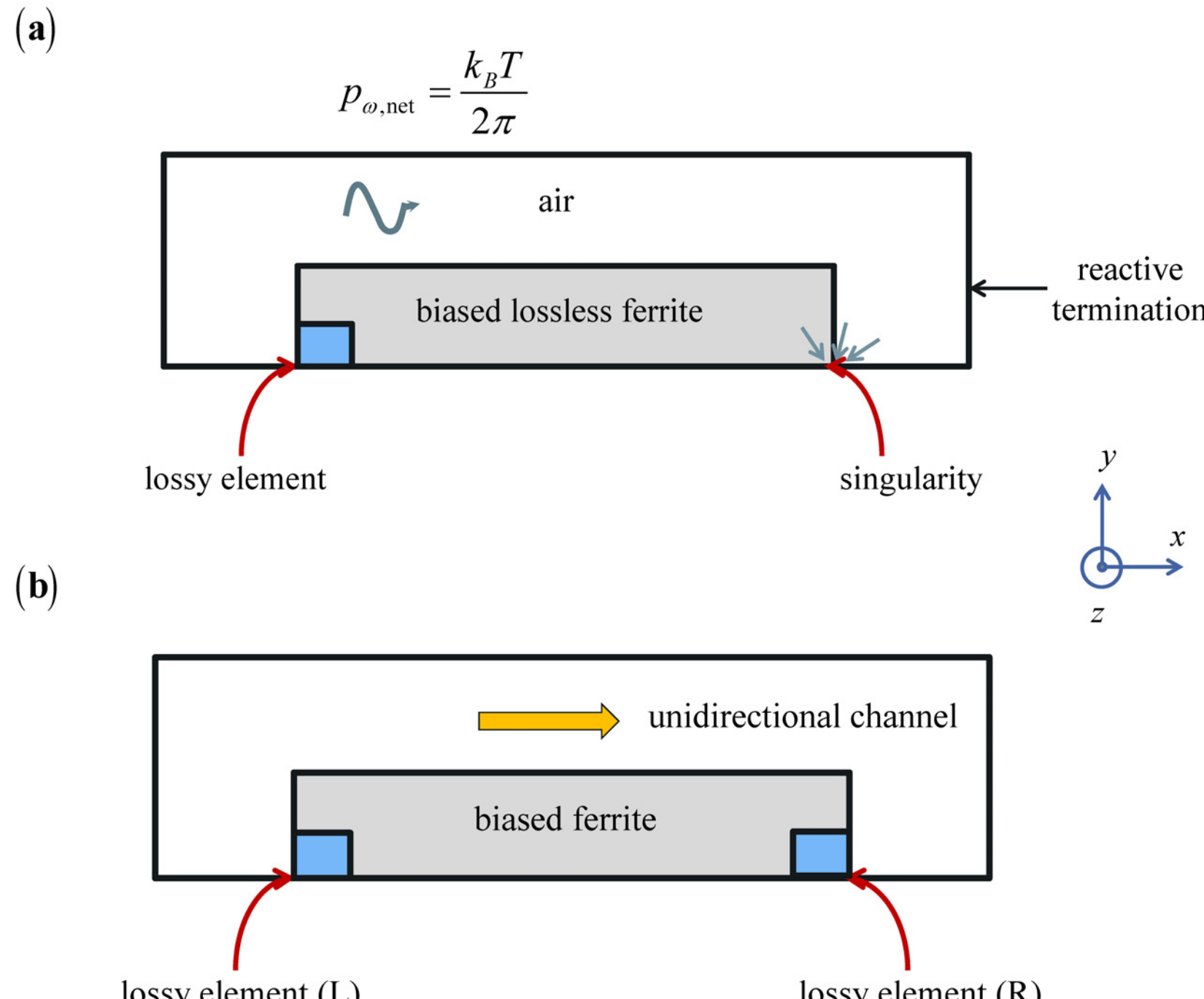

**Fig. 1** **(a)** Sketch of a metallic cavity partially filled with a magnetized ferrite and terminated by reactive walls. It is assumed that, in the relevant spectral band, the waveguide section supports a single propagating mode, directed toward the right. Material loss is considered negligible everywhere except in the blue-shaded element on the left-hand side. Due to the absence of dissipative elements elsewhere, the energy transported through the guide accumulates at a hotspot, where the fields exhibit singular behavior. **(b)** Geometry similar to (a), but with dissipative elements at both the left and right ends of the ferrite slab. The dissipative "R" element prevents energy accumulation at the system's right end, but does not resolve the underlying thermodynamic paradox.

To illustrate the unidirectional behavior of this system, Fig. 2 shows the dispersion relation (green curve) of the fundamental mode of the relevant rectangular metallic waveguide filled with a ferrite slab of thickness *d*. The cross-section of the metallic waveguide is $a \times b$ with $a$ the largest dimension. The fundamental mode is characterized by fields of the form $\mathbf{E} = E_z \hat{\mathbf{z}}$ and $\mathbf{H} = H_x \hat{\mathbf{x}} + H_y \hat{\mathbf{y}}$, corresponding to a transverse electric (TE) polarization. A static magnetic bias is applied along the +$z$-direction, giving the ferrite a gyrotropic magnetic response of the type [1]:

$$\overline{\mu} = \mu_0 \begin{pmatrix} \mu_t & -i\mu_g & 0 \\ i\mu_g & \mu_t & 0 \\ 0 & 0 & 1 \end{pmatrix}, \tag{1a}$$

$$\mu_t = 1 + \frac{A\omega_0\omega_{\mathrm{m}}}{\omega_0^2 A^2 - \omega^2}, \qquad \mu_g = \frac{\omega\omega_{\mathrm{m}}}{\omega_0^2 A^2 - \omega^2}, \tag{1b}$$

with $A = 1 - i\alpha\,\omega/|\omega_0|$. Here, $\omega_0$ is the (Larmor) cyclotron frequency proportional to the applied magnetic bias, $\omega_{\mathrm{m}}$ is the Larmor frequency for the saturation magnetization, and $\alpha > 0$ controls the strength of the damping due to material absorption. The electric response of the ferrite is characterized by a scalar permittivity $\varepsilon_0\varepsilon_r$. The dispersion of the guided mode is obtained using the theory in Ref. [1, Sect. 9.3].

As shown in Fig. 2 for typical waveguide parameters, there exists a broad spectral band in which the structure supports a single forward-propagating mode (along +$x$). Importantly, this unidirectional behavior is robust: regardless of the precise design parameters, the waveguide always supports a purely forward mode within the spectral window

$$\left|\omega_0 + \frac{\omega_{\mathrm{m}}}{2}\right| \leq \omega \leq \left|\omega_0 + \omega_{\mathrm{m}}\right|, \tag{2}$$

which, in the example of Fig. 2, corresponds to the interval $1.0 \leq \omega/\omega_{\mathrm{m}} \leq 1.5$.

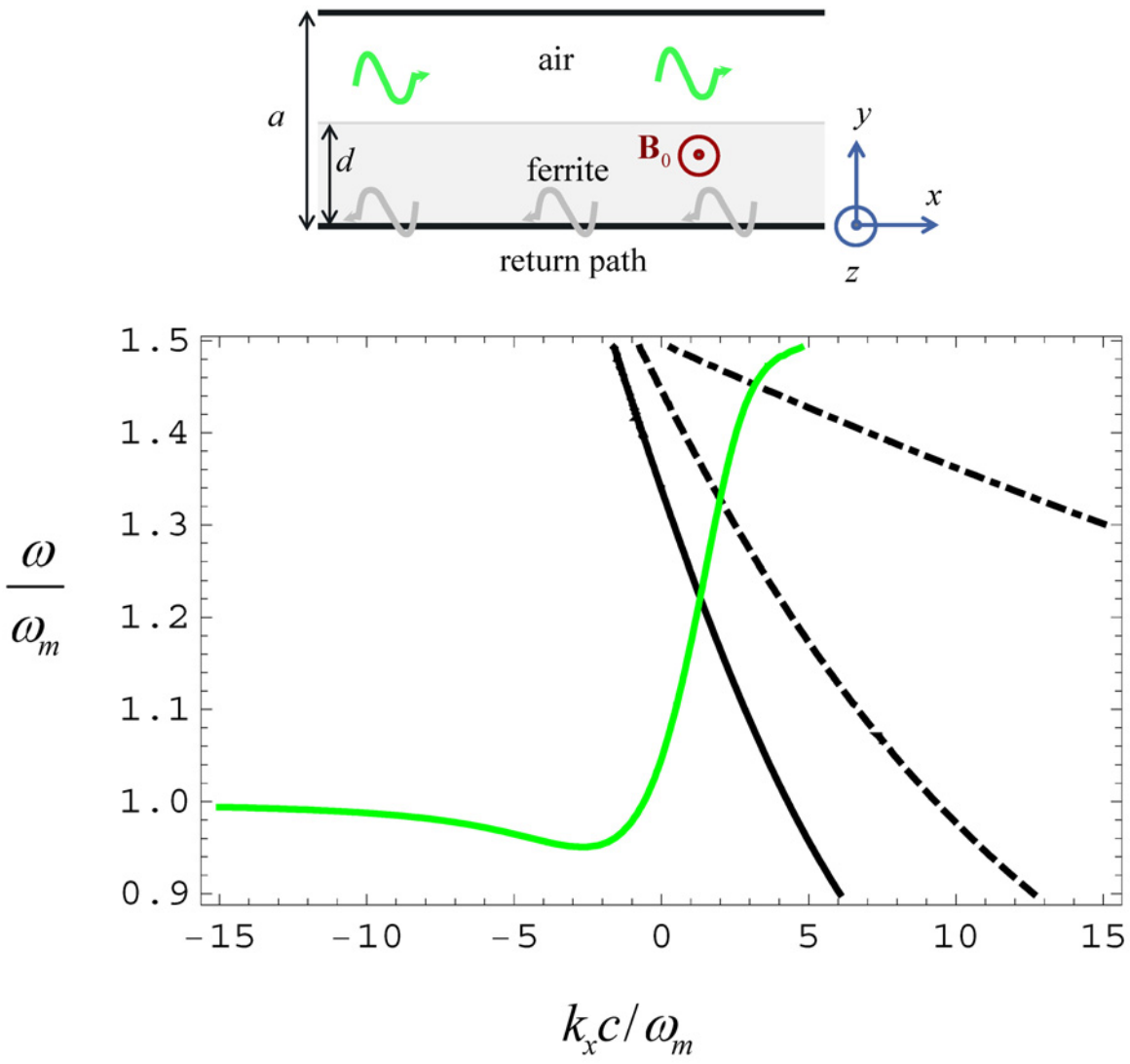

**Fig. 2** Dispersion $\omega(k_x)$ of the fundamental mode of a rectangular waveguide with cross-section $a \times b$ partially filled with a magnetized ferrite. The ferrite parameters are $\varepsilon_r = 10$, $\omega_0 = 0.5\omega_m$ with $\omega_m = c/a$, with a slab thickness $d = 0.5a$. The inset shows a cross-sectional view of the waveguide in the *xoy* plane. The direction of propagation is along *x*. Green line: dispersion calculated using the local model of the ferrite. Black lines: additional guided wave that appears when a wave vector cut-off is imposed: $k_{\max} = 10/a$ (solid black line), $k_{\max} = 20/a$ (dashed black line), $k_{\max} = 100/a$ (dot-dashed black line). This additional guided wave (gray arrows in the inset) propagates strongly bound to the bottom plate and serves as a return path for the forward-propagating mode (green arrows) when the waveguide is terminated by a reactive load.

The paradox arising from closing the waveguide ends with reactive walls is straightforward to describe. As we will discuss in detail in Sect. II, thermal radiation originates from the noise currents within the materials, whose strength is directly tied to the material's dissipative properties. Consider the waveguide of Fig. 1a operating in a spectral band where propagation is strictly unidirectional, and suppose that all the material loss is confined to a single element (shaded in blue). According to Nyquist–Johnson theory [12], and more generally the fluctuation–dissipation theorem [13, 14], the spectral density of power transported by a waveguide mode in the classical limit ($\hbar\omega << k_B T$) is:

$$p_\omega\big|_{\text{mode}} = \frac{k_B T}{2\pi} s\,, \tag{3}$$

where $k_B$ is the Boltzmann constant and $T$ is the temperature. In this expression, $s$ indicates the direction of propagation of the mode, with $s$=+1 corresponding to power flowing to the right, and $s$=−1 to power flowing to the left. This result assumes that the waveguide exhibits negligible dissipation except near the terminations.

When the waveguide is strictly unidirectional, supporting a single propagating mode to the right, the thermal energy emitted by the dissipative element at the left end couples into that mode, as it is the only available radiative channel in the relevant spectral region. However, the emitted radiation then propagates to the right end, which is terminated with a reactive wall. If material dissipation in the guide and termination is neglected a problem arises: the thermal energy generated at the left-hand side travels to the right termination, where there is no mechanism for it to be dissipated or reflected back: no available radiative channels exist to return it, nor are there any absorptive channels to remove it from the system. As a result, energy accumulates at the right-end termination, which is clearly incompatible with a stationary (thermodynamic) equilibrium.

A widely held view is that material dissipation provides the essential resolution: it allows the thermal energy emitted from the left-hand side to be absorbed at the right-hand side, thereby avoiding unphysical singular behaviors [10]. This interpretation also appears consistent with numerical studies of externally driven systems, which show that adding dissipation at the waveguide right-end enables the formation of a steady state [15-20]. While this reasoning captures part of the physics, we argue that loss alone is not sufficient to resolve the paradox. The crucial point is that, in true thermodynamic equilibrium, the power emitted by any material element must be exactly compensated by the power it absorbs.

To make this concrete, consider the idealized system sketched in Fig. 1b: the same waveguide, but now with both ends terminated by dissipative elements, while the central region remains lossless. Introducing loss at both terminations does prevent the unphysical energy build-up at the right-hand end [10], as discussed earlier. However, true thermal equilibrium requires more than simply removing energy: it requires that the power transferred from the left dissipative element to the right one be exactly equal to the power transferred in the reverse direction:

$$p_{\omega,\mathrm{L}\to\mathrm{R}} = p_{\omega,\mathrm{R}\to\mathrm{L}}\,. \tag{4}$$

In Sect. II, we will show explicitly that this intuitive condition follows from the fluctuation–dissipation theorem. Yet if the central region of the waveguide is perfectly lossless, the right-hand (“R”) element is effectively isolated from the left-hand (“L”) element: no channel exists for returning the energy absorbed by “R” back to “L”. Thus, the inclusion of dissipation at both ends does not, by itself, resolve the paradox.

One might object that confining dissipation to two isolated regions is artificial, since real materials exhibit distributed loss. When absorption is present throughout the waveguide, there is no strict separation between the “L” and “R” regions; any partition inevitably couples through reactive fields, as the two sides lie within each other’s near field. Nevertheless, as we shall demonstrate in the following sections, even when loss is distributed across the system, the thermodynamic paradox persists, manifesting itself through non-integrable eigenmode singularities and delta-type contributions to the heat current.

This article is organized as follows. In Sect. II we review the fluctuation–dissipation theorem and derive a modal expansion for the heat current in dispersive dissipative systems. In Sect. III we show that, in the problematic spectral region, cavity eigenmodes exhibit non-integrable singularities even under strong dissipation. In Sect. IV we demonstrate that nonlocality provides a natural resolution by creating a topological return channel. We link the removal of singularities with well-defined topology, even in the presence of dissipation. We further establish a weak form of bulk–edge correspondence in passive non-Hermitian systems, grounded in the requirement of thermal equilibrium. Section V summarizes the main findings and implications.

## II. Thermal light

### *A. Fluctuation-dissipation relations*

Thermal light is electromagnetic radiation emitted by matter characterized by a broad incoherent spectrum determined by the statistical fluctuations of the material's microscopic charges. These fluctuations are described by noise currents, whose unilateral spectral correlations are governed by generalized Nyquist–Johnson formulas [12, 21, 22]:

$$\left\langle \left\{ \mathbf{j}_{\mathrm{N}}\left(\mathbf{r}\right)\mathbf{j}_{\mathrm{N}}\left(\mathbf{r}'\right)\right\} \right\rangle_{T,\omega} = \hbar\omega N_{\omega}\delta\left(\mathbf{r}-\mathbf{r}'\right)\frac{2}{\pi}\mathrm{Re}\left\{ \omega\mathbf{M}''\right\}. \tag{5}$$

Here, $N_{\omega} = \dfrac{1}{e^{+\beta\hbar\omega}-1}$ is the occupation number for a quantum harmonic oscillator with frequency $\omega$ and $\beta = 1/k_B T$ is the inverse temperature. The equivalent noise currents at different spatial points are uncorrelated. The strength of the current–current correlations is directly determined by the dissipative properties of the medium through the matrix $\mathbf{M}'' = \frac{\mathbf{M} - \mathbf{M}^{\dagger}}{2i}$. For convenience, we adopt a 6-vector notation such that $\mathbf{j}_{\mathrm{N}}(\mathbf{r})$ is a vector that includes the equivalent electric and magnetic noise currents $\mathbf{j}_{\mathrm{N}}(\mathbf{r}) = \begin{pmatrix} \mathbf{j}_{\mathrm{e,N}}(\mathbf{r}) & \mathbf{j}_{\mathrm{m,N}}(\mathbf{r}) \end{pmatrix}^{T}$.

In its most general form, the material matrix $\mathbf{M} = \mathbf{M}(\mathbf{r}, \omega)$ links the $\mathbf{D}, \mathbf{B}$ macroscopic fields with the $\mathbf{E}, \mathbf{H}$ fields as follows:

$$\begin{pmatrix} \mathbf{D} \\ \mathbf{B} \end{pmatrix} = \mathbf{M} \cdot \begin{pmatrix} \mathbf{E} \\ \mathbf{H} \end{pmatrix} \equiv \begin{pmatrix} \varepsilon_0 \overline{\varepsilon} & \frac{1}{c} \overline{\xi} \\ \frac{1}{c} \overline{\zeta} & \mu_0 \overline{\mu} \end{pmatrix} \cdot \begin{pmatrix} \mathbf{E} \\ \mathbf{H} \end{pmatrix}. \tag{6}$$

As usual, $\overline{\varepsilon}$ represents the frequency dependent permittivity tensor, $\overline{\mu}$ is the permeability tensor, and $\overline{\xi}, \overline{\zeta}$ describe magneto-electric interactions. In this article, we focus on (non-bianisotropic) materials that exhibit a nonreciprocal magnetic response.

The correlations of the electromagnetic field can be determined by inserting the current–current correlations into Maxwell's equations. These correlations are governed by the fluctuation-dissipation theorem, which states that, in a general electromagnetic environment, the one-sided spectral density of the field correlations is given by [21-22]:

$$\left\langle \left\{ \mathbf{f}(\mathbf{r}) \mathbf{f}(\mathbf{r}') \right\} \right\rangle_{T,\omega} = N_{\omega} \operatorname{Re} \left\{ \frac{\hbar}{i\pi} \left( \overline{\mathcal{G}}(\mathbf{r}, \mathbf{r}', \omega) - \left[ \overline{\mathcal{G}}(\mathbf{r}', \mathbf{r}, \omega) \right]^{\dagger} \right) \right\}. \tag{7}$$

Here, $\mathbf{f}(\mathbf{r}) = \begin{pmatrix} \mathbf{E}(\mathbf{r}) & \mathbf{H}(\mathbf{r}) \end{pmatrix}^{T}$ is a six-vector that includes both the electric and magnetic fields. Furthermore, $\overline{\mathcal{G}}(\mathbf{r}, \mathbf{r}', \omega)$ is the system Green's function defined as the solution of the Maxwell's equations for a generic dipole-type excitation:

$$\hat{\mathcal{D}} \cdot \overline{\mathcal{G}} - \omega \mathbf{M}(\mathbf{r}, \omega) \cdot \overline{\mathcal{G}} = \omega \mathbf{1} \delta(\mathbf{r} - \mathbf{r}'). \tag{8}$$

In the above, $\mathbf{r}$ is the observation point, $\mathbf{r}'$ is a generic source point. The differential operator $\hat{\mathcal{D}}$ is defined as

$$\hat{\mathcal{D}} = \begin{pmatrix} \mathbf{0}_{3\times3} & i\nabla \times \mathbf{1}_{3\times3} \\ -i\nabla \times \mathbf{1}_{3\times3} & \mathbf{0}_{3\times3} \end{pmatrix}. \tag{9}$$

We are primarily interested in characterizing the flow of thermal energy in a closed cavity (Fig. 1), specifically through the expectation value of the Poynting vector associated with thermal radiation, commonly referred to as the heat current [23-24]. Intuitively, in thermodynamic equilibrium, the expectation of the Poynting vector should vanish, as this is the simplest way to ensure that there is no net absorption or emission by any generic material element. This is indeed what occurs in reciprocal systems, which are characterized by a trivial heat current with a trivial spectrum: $\mathbf{S}_\omega(\mathbf{r}) = 0$ [23-24].

However, in the general nonreciprocal case, this condition is no longer mandatory, and a nontrivial heat current can still be fully compatible with thermodynamic equilibrium [23-25]. In fact, maintaining equilibrium, i.e., ensuring that the net absorption by each material element is exactly balanced by its net emission only requires the absence of sources and sinks of thermal radiation. Mathematically this condition is expressed as:

$$\nabla \cdot \mathbf{S}_\omega(\mathbf{r}) = 0. \tag{10}$$

In other words, thermodynamic equilibrium requires that the divergence of the heat current vanish, ensuring that the Poynting vector field lines associated with thermal radiation form closed orbits. This constraint implies that the power exchanged between any partition of the noise currents (let us say, between left and right elements) must be identical, as expressed by Eq. (4).

### *B. Modal expansion of the heat current*

In the following, we express the heat current expectation in terms of the normal modes of the system. This can be achieved through a modal expansion of the system's Green's function. Since realistic systems are dissipative, i.e., they exhibit a non-Hermitian response, the expansion is carried out using a bi-orthogonal basis of the system [26, 27]. Bi-orthogonal expansions involve not only the eigenmodes of the original system but also those of the Hermitian-conjugate problem.

To develop this idea further, let $\mathbf{f}_n(\mathbf{r};\alpha) = (\mathbf{E}_n \quad \mathbf{H}_n)^T$ be a family of eigenmodes of the system with corresponding eigenfrequencies $\omega_n(\alpha)$. For passive systems, these eigenfrequencies must lie in the lower half of the complex frequency plane. The eigenmodes are indexed by a real-valued parameter $\alpha$, which controls the strength of the dissipative response (see Eq. (1)). More generally, $\alpha$ can be viewed as a set of parameters governing the system's dissipative properties, for instance, the absorption rates (e.g., collision frequencies) of the constituent materials. In the limit $\alpha = 0$, the system response becomes conservative.

The Hermitian-conjugate problem corresponds to a system with symmetric decay rates, i.e., where absorption is replaced by gain, which is equivalent to flipping the sign of $\alpha$. Accordingly, the eigenmodes of the Hermitian-conjugate system take the form $\mathbf{f}_n^c \equiv \mathbf{f}_n(\mathbf{r};-\alpha)$, and they are associated with eigenfrequencies satisfying $\omega_n^c = \omega_n(-\alpha) = \omega_n^*(\alpha)$ [26]. As a result, the eigenfrequencies of the Hermitian-conjugate problem lie in the upper-half complex frequency plane.

In Appendices A, B, and C, we develop a comprehensive theory for bi-orthogonal modal expansions in dissipative (passive) dispersive systems. Our framework generalizes earlier results that focused on conservative systems [28, 29]. In particular, we show that the system Green's function [Eq. (8)] can be written in terms of the normal modes as follows [Eq. (B12)]

$$\overline{\mathcal{G}}(\mathbf{r},\mathbf{r}',\omega) = \frac{\omega}{2}\sum_n \frac{1}{\omega_n - \omega}\mathbf{f}_n(\mathbf{r};\alpha)\otimes\mathbf{f}_n^*(\mathbf{r}';-\alpha). \tag{11}$$

The eigenmodes are normalized according to Eq. (B11) and are assumed to form a complete set. Substituting this expression into the fluctuation-dissipation relation [Eq. (7)], we find that it can be recast as:

$$\begin{aligned}&\left\langle\left\{\mathbf{f}(\mathbf{r})\mathbf{f}(\mathbf{r}')\right\}\right\rangle_{T,\omega}\\ &= \hbar\omega N_\omega \operatorname{Re}\left\{\frac{i}{\pi}\sum_n \frac{1}{\omega-\omega_n}\frac{1}{2}\left[\mathbf{f}_n(\mathbf{r};\alpha)\otimes\mathbf{f}_n^*(\mathbf{r}';-\alpha)+\mathbf{f}_n^*(\mathbf{r};-\alpha)\otimes\mathbf{f}_n(\mathbf{r};\alpha)\right]\right\}\end{aligned} \tag{12}$$

In particular, the (unilateral) Poynting vector spectrum is determined by:

$$\mathbf{S}_\omega(\mathbf{r}) = \hbar\omega N_\omega \operatorname{Re}\left\{\frac{i}{\pi}\sum_n \frac{1}{\omega-\omega_n}\mathbf{S}_n(\mathbf{r})\right\}, \qquad \text{with} \tag{13a}$$

$$\mathbf{S}_n(\mathbf{r};\alpha) \equiv \frac{1}{2}\left[\mathbf{E}_n(\mathbf{r};\alpha)\times\mathbf{H}_n^*(\mathbf{r};-\alpha)+\mathbf{E}_n^*(\mathbf{r};-\alpha)\times\mathbf{H}_n(\mathbf{r};\alpha)\right]. \tag{13b}$$

This formula generalizes the result of Ref. [24], $\mathbf{S}_\omega(\mathbf{r}) = \hbar\omega N_\omega \sum_n \delta(\omega-\omega_n)\mathbf{S}_n(\mathbf{r})$, which holds in the Hermitian case. The proof follows directly from the Sokhotski–Plemelj identity and from the fact that, in the limit $\alpha \to 0^+$, both the eigenfrequencies $\omega_n$ and $\mathbf{S}_n$ are real valued. Furthermore, we note that a modal expansion similar to Eq. (13) was recently reported in Ref. [30], although it did not explicitly account for the effects of material dispersion.

For simplicity, and without loss of generality, in the remainder of the article we focus on the spectral region $\beta\hbar\omega << 1$, corresponding to the classical limit. In this regime, one can use the approximation $N_\omega \approx \frac{k_B T}{\hbar\omega}$, which leads to the simplified expression for the heat current spectrum:

$$\mathbf{S}_\omega(\mathbf{r}) \approx k_B T \operatorname{Re}\left\{ \frac{i}{\pi} \sum_n \frac{1}{\omega - \omega_n} \mathbf{S}_n(\mathbf{r};\alpha) \right\}. \tag{14}$$

In Appendix D, we demonstrate explicitly that the above formula is compatible with the conservation law given in Eq. (10), i.e., it ensures the heat current expectation is free of sinks and sources, as required in thermodynamic equilibrium (see also Ref. [30]). The proof implicitly assumes that the eigenmodes are free of singularities. However, as we will show in the following section, this assumption can fail in regions of transition between reciprocal and nonreciprocal materials, leading to problematic equilibrium behavior.

## III. Ill-posed Electromagnetic Behavior

Next, we show that the true origin of the thermodynamic paradox is rooted in the ill-posedness of Maxwell's equations in certain nonreciprocal environments. Specifically, we demonstrate that in geometries involving nonreciprocal materials, the electromagnetic modes may fail to be square-integrable, even in the presence of strong material dissipation. We show that this divergent behavior effectively introduces sinks and sources of thermal energy near the cavity boundaries, thereby violating the conditions required for a true thermodynamic equilibrium.

### *A. Singularities of the normal modes*

To illustrate how certain transitions between reciprocal and nonreciprocal elements can give rise to non-integrable field singularities, we first focus on the "R" region in Fig. 1b). This region can be modelled as a 180º metallic wedge (a metallic plate) half-filled with magnetized ferrite and air, as shown in detail in Fig. 3a). It is well known that electromagnetic fields near sharp wedges can exhibit singular behaviour [31]. However, in our case, the metallic wedge is as smooth as possible: it is, in fact, a flat metallic plate. Thus, any singular behaviour that arises in this configuration cannot be attributed to the metal geometry itself.

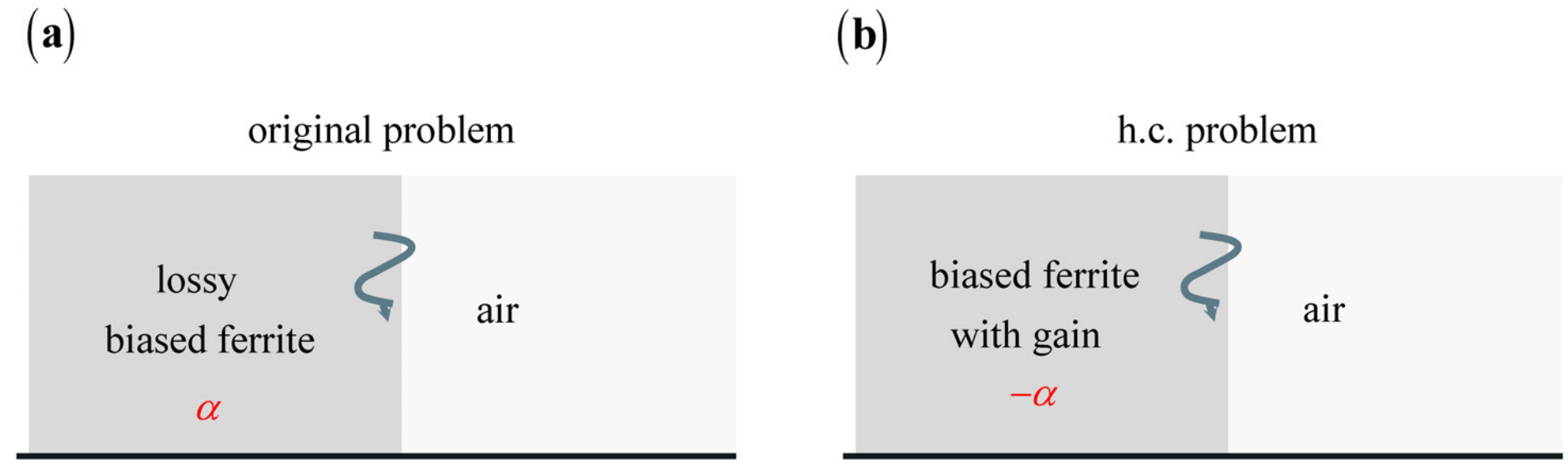

**Fig. 3 (a)** Detail of the "R" region in Fig. 1b), corresponding to an 180º degree metallic wedge (metallic plate) half-filled with a lossy magnetized ferrite and air. **(b)** Geometry of the Hermitian-conjugate problem, in which the lossy ferrite is replaced by a ferrite exhibiting a gain response.

We characterize the field behavior near the wedge using standard methods, adopting a quasi-static approximation that neglects retardation effects (i.e., assumes an infinite speed of light $c \to \infty$) [10, 15, 31]. Within this framework, the electric field in the vicinity of the wedge can be expressed as $E_z = \rho^{\nu} g(\varphi)$, where $(\rho, \varphi, z)$ are cylindrical coordinates referenced to the wedge vertex (i.e. the point where the three material regions meet). The exponent $\nu$ governs the asymptotic behavior of the electromagnetic fields as the distance to the vertex $\rho$ approaches zero, while the function $g$ specifies the azimuthal variation. The details of the formulation are presented in Appendix E. We find that the allowed exponents $\nu$ are in general a function of frequency and the material response.

We begin with the simple case in which the ferrite is not magnetized, so that its response is reciprocal. In this situation, regardless of the ferrite's dispersive properties or material loss, the allowed exponents are $\nu = \pm n$ with $n = 1, 2, 3, \ldots$. The negative exponents must be discarded, as they correspond to strongly singular behaviors that cannot be excited by any physical source located away from the vertex. Therefore, realistic field distributions correspond to solutions with $\nu = 1, 2, 3, \ldots$, which describe an electric field that vanishes at the wedge vertex, consistent with the perfect electric conductor (PEC) boundary condition. In particular, for a reciprocal system, the electromagnetic fields near a 180º wedge do not exhibit any singular behavior, irrespective of the ferrite's dispersion (for TE polarized fields, as in Fig. 1).

When a static magnetic field is applied, the previous discussion changes substantially. Specifically, while the even-valued exponents $\nu = \pm 2, \pm 4, \ldots$ remain independent of the ferrite's material dispersion, the odd-valued exponents $\nu = \pm 1, \pm 3, \pm 5, \ldots$ become sensitive to the bias and frequency dependent. Since the even-valued exponents play no role in our problem, they will be disregarded in the following.

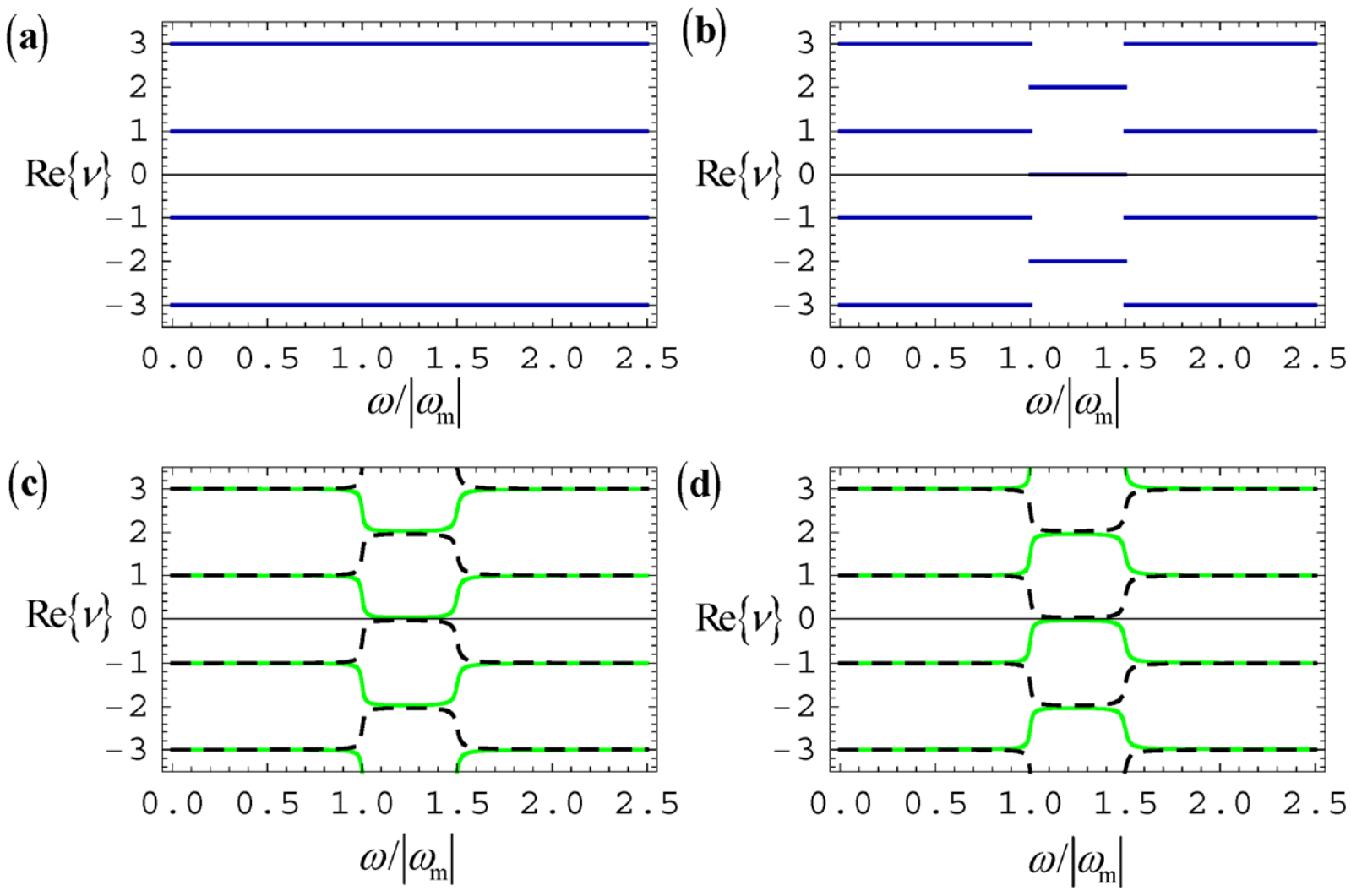

**Fig. 4** Plot of the real part of the singularity exponent $\nu$ as a function of normalized frequency for a wedge with the geometry of Fig. 3. **(a)** Reciprocal case. **(b)** Lossless magnetized ferrite. **(c)** Magnetized ferrite with a damping factor $\alpha = 0.01$ and bias magnetic field oriented along $+z$ ($\omega_{\mathrm{m}} > 0$). Green lines: dissipative system; Dashed black lines: Hermitian conjugate problem. **(d)** Similar to c) but with the bias magnetic field oriented along $-z$ ($\omega_{\mathrm{m}} < 0$). In all panels, the cyclotron frequency is $\omega_0 = 0.5\omega_{\mathrm{m}}$.

Importantly, in the lossless case ($\alpha = 0$), the allowed exponents $\nu$ exhibit a discontinuous variation with frequency, even for an arbitrarily weak magnetic bias ($\omega_0, \omega_{\mathrm{m}} \to 0$). This behavior is illustrated in Fig. 4b), which shows the real part of $\nu$ as a function of normalized frequency for the same material parameters as in Fig. 2. The curve is obtained from [Eq. (E9) in Appendix E]

$$\cot\left(\frac{\pi}{2}\nu\right) = \frac{-i\mu_g}{\mu_t^2 - \mu_g^2 + \mu_t}, \tag{15}$$

which determines the admissible values of $\nu$. Note that $\nu$ is a multi-valued function with different branches separated by an even integer number. For comparison, Fig. 4a shows the branches of $\nu$ when the magnetic bias is removed, highlighting how the nonreciprocal response can dramatically alter the behavior of the fields in the vicinity of the vertex point.

As seen in Fig. 4b, with a magnetic bias there exists a frequency window in which one branch of $\nu$ has a vanishing real part. Remarkably, even though the calculation of $\nu$ relies on a quasi-static approximation the frequency window where $\mathrm{Re}\{\nu\}=0$ with $\mathrm{Im}\{\nu\}\neq 0$ is exactly coincident with the spectral window where the corresponding waveguide is unidirectional [Eq. (2)]. As demonstrated in Appendix E, solutions with $\mathrm{Re}\{\nu\}\leq 0$ produce strongly singular, non-square-integrable field distributions because the electromagnetic fields have the following asymptotic behavior near the wedge vertex: $E_z \sim \rho^{\nu}$ and $\mathbf{H} \sim \rho^{\nu-1}$.

Incorporating dissipation into the ferrite response removes the discontinuous transitions in the branches of $\nu$. This effect is illustrated by the green solid lines in Fig. 4c), computed with the damping factor $\alpha = 0.01$, assuming a magnetic bias oriented along +*z*-direction. As shown, in the presence of dissipation the branch with $\mathrm{Re}\{\nu\}=0$ is shifted into the region where the solution becomes physically admissible and the fields are square-integrable ($\mathrm{Re}\{\nu\}>0$).

This observation is consistent with the earlier work of Ishimaru, who showed that dissipation alone could eliminate the unphysical singularity at point "R" and restore the well-posedness of the electromagnetic problem [10] (see also Ref. [15]). Indeed, with any finite dissipation level, the electromagnetic energy stored near the vertex becomes finite, ensuring a well-defined steady state under time-harmonic excitation.

It is worth noting that dissipation alone is not sufficient to render the magnetic field finite near the wedge, albeit it becomes square-integrable. This persistent singular behavior is associated with the formation of hotspots, as is well documented in both numerical [15-20] and experimental studies [11, 32]. In particular, in a notable experiment [11], Barzilai and Gerosa visually demonstrated the formation of a hotspot near point "R" in Fig. 1b). In their experiment, the waveguide was coated with a temperature-sensitive dye and excited with bursts of microwave radiation. The authors observed a distinct hotspot forming at point "R", marking the transition between the propagating region of the waveguide and the reactive termination. This experimental evidence clearly confirms the physical relevance of the branch with $\mathrm{Re}\{\nu\} \approx 0^+$.

It is relevant to examine how the field behavior near the wedge changes when the material response exhibits gain rather than dissipation, corresponding to the Hermitian-conjugate problem illustrated in Fig. 3b). The effect of gain is modeled by a negative damping factor $\alpha = -0.01$. The resulting branches of the singularity exponent $\nu$ are shown in Fig. 4c) as dashed black lines. In contrast to the dissipative case, gain shifts the branch with $\mathrm{Re}\{\nu\} = 0$ into the range of exponents with $\mathrm{Re}\{\nu\} < 0$. Intuitively, this behavior is consistent with physical expectations: dissipation tends to suppress singularities near the point where the guided wave is abruptly halted, by attenuating the mode; conversely, gain tends to enhance singular behavior by amplifying the wave. It is worth noting that for a vanishingly small $|\alpha|$, the results for both lossy and gain scenarios converge to those of the ill-posed Hermitian problem, which exhibits spectral regions with $\mathrm{Re}\{\nu\} = 0$, as shown in Fig. 4b).

### *B. Heat current sink*

So far, our discussion has been limited to the case of real-valued frequencies. However, as already discussed in Sec. III.B, the normal modes of the system are, in general, associated with complex-valued frequencies. In particular, the heat current is expressed in terms of these normal modes with complex frequencies [Eq. (13)].

The behavior of the exponent $\nu$ for complex frequencies can be analyzed using the same methods as in the real-frequency case. In particular, Eq. (15) still governs the variation of $\nu = \nu(\omega;\alpha)$. In Appendix E, we show that the singularity exponents of the original problem with dissipation are generally related to those of the Hermitian-conjugate problem with gain as follows:

$$\nu(\omega^*;-\alpha) = -\nu^*(\omega;\alpha). \tag{16}$$

As shown next, this result allows us to connect the singularity exponents of partner modes of the system bi-orthogonal basis [see Appendices A and B for further discussion of the bi-orthogonal basis].

Indeed, let $\nu_n$ ($\nu_n^{\mathrm{c}}$) represents the singularity exponents for the normal modes $\mathbf{f}_n(\mathbf{r};\alpha)$ ($\mathbf{f}_n(\mathbf{r};-\alpha)$) of the original (Hermitian-conjugate) problem, respectively. The partner modes $\mathbf{f}_n(\mathbf{r};\alpha)$ and $\mathbf{f}_n(\mathbf{r};-\alpha)$ are associated with complex-conjugated eigenfrequencies [Eq.(B1)]. As previously noted, the function $\nu = \nu(\omega;\alpha)$ is multi-valued, with its branches separated by even integers. Combining Eq. (16) with this property, we find that the singularity exponents of partner modes are related by $\nu_n^{\mathrm{c}} = -\nu_n^* + 2m$, for some integer $m$. The integer $m$ can be found noting that, in the limit of vanishingly small damping ($|\alpha| \to 0$), the exponents $\nu_n^{\mathrm{c}}, \nu_n$ must coincide, as the problem becomes Hermitian.

For normal modes with eigenfrequencies close to the problematic spectral range, $\left|\omega_0 + \frac{\omega_{\mathrm{m}}}{2}\right| \le \omega \le |\omega_0 + \omega_{\mathrm{m}}|$, where $\mathrm{Re}\{\nu\} \approx 0$ [see Eq. (2)] the collapse of the two solutions occurs for $m = 0$. Thereby, we have that:

$$\mathrm{Re}\left\{\nu_n^{\mathrm{c}}\right\} = -\mathrm{Re}\left\{\nu_n\right\}, \qquad \text{(in the singular spectral range).} \tag{17}$$

This property is central to revealing the thermodynamic paradox.

Indeed, the “mixed” Poynting vector defined in Eq. (13b), involves the product of a field of the original problem and a field of the Hermitian conjugate problem. Since the asymptotic behavior of the fields is governed by $\mathbf{E}_n(\mathbf{r};\alpha) \sim \rho^{\nu_n}$, $\mathbf{H}_n(\mathbf{r};\alpha) \sim \rho^{\nu_n - 1}$, $\mathbf{E}_n(\mathbf{r};-\alpha) \sim \rho^{\nu_n^{\mathrm{c}}}$, $\mathbf{H}_n(\mathbf{r};\alpha) \sim \rho^{\nu_n^{\mathrm{c}} - 1}$, it is straightforward to verify that the asymptotic behavior of the mixed Poynting vector [Eq. (13b)] in the problematic spectral range is governed by

$$\mathbf{S}_n(\mathbf{r};\alpha) \sim \rho^{\mathrm{Re}\nu_n + \mathrm{Re}\nu_n^{\mathrm{c}} - 1} = \frac{1}{\rho}, \tag{18}$$

near the wedge vertex. The rightmost identity follows from Eq. (17). Thus, the tamed singularity in the original system (due to the presence of dissipation) is exactly compensated by the enhanced singularity in the Hermitian-conjugate problem (due to the presence of gain), so that the mixed Poynting vector retains a singular behavior that is entirely independent of the damping coefficient $\alpha$.

Evidently, the heat current spectrum in Eq. (13a) exhibits the same singular behavior near the wedge vertex, $\mathbf{S}_\omega(\mathbf{r}) \sim 1/\rho$. In particular, the divergence of the heat current near the wedge vertex displays a delta-type (line) singularity, $\nabla \cdot \mathbf{S}_\omega(\mathbf{r}) \sim \delta(x)\delta(y)$, indicative of a localized heat sink that is incompatible with thermodynamic equilibrium. Thus, the paradox persists even in the presence of strong material dissipation. It arises from the singular behavior of the fields in the vicinity of the wedge vertex, and therefore persists regardless of whether material loss is distributed throughout the entire ferrite or confined only to the terminations.

In the present geometry, the singular point ("R") lies on the boundary of the cavity domain. However, for other configurations, such a singularity could occur within the interior of the cavity. Moreover, although our analysis here assumes an abrupt transition between the magnetized ferrite and the air region, we expect a similar singular behavior if the ferrite parameters vary smoothly across a transition layer connecting the air and ferrite responses. In fact, a smooth variation of material parameters does not necessarily eliminate such singularities, as demonstrated in Ref. [33] for a related platform.

In summary, the picture that emerges from the preceding discussion is that dissipation regularizes the electromagnetic response near the hotspot associated with point "R", ensuring a finite stored energy and a well-defined steady state under time-harmonic excitation. Conversely, gain amplifies the singularity at point "R" relative to the Hermitian case. Since the heat current is governed by a self-term of the Green's function, $\mathbf{S}_\omega(\mathbf{r}) \sim \overline{\mathcal{G}}(\mathbf{r},\mathbf{r},\omega) - \left[\overline{\mathcal{G}}(\mathbf{r},\mathbf{r},\omega)\right]^\dagger$, the two effects cancel, leading to the formation of a heat sink.

At first glance, this picture might suggest that the problematic behavior could occur only for fluctuation-induced fields, since the fields generated by an external time-harmonic excitation would not be affected by the self-field away from the source point.

In the next subsection, we demonstrate that this is not the case: the presence of a singularity in the Hermitian-conjugate problem inevitably entails the existence of a corresponding singularity somewhere within the original dissipative cavity.

### *C. Reciprocity and heat current source*

Although the Lorentz reciprocity theorem does not directly apply to a nonreciprocal platform, it is possible to relate the modes of the reciprocal dual system to those of the Hermitian-conjugate system, as shown next. The reciprocal dual is defined as a system in which all parameters governing the macroscopic response that are odd under time reversal are inverted. Formally, the material response of the reciprocal dual is given by

$$\mathbf{M}_{\mathrm{rd}} = \boldsymbol{\sigma} \cdot \mathbf{M}^T \cdot \boldsymbol{\sigma}, \quad \text{with} \quad \boldsymbol{\sigma} = \begin{pmatrix} \mathbf{1}_{3\times3} & 0 \\ 0 & -\mathbf{1}_{3\times3} \end{pmatrix}, \tag{19}$$

where superscript $T$ denotes the matrix transpose. Here, $\mathbf{M}_{\mathrm{rd}}$ denotes the response of the reciprocal dual system, while $\mathbf{M}$ corresponds to the material matrix of the original system.

In the specific system studied in this work, the reciprocal dual corresponds to a waveguide in which the magnetic bias is reversed, $\mathbf{B}_0 \to -\mathbf{B}_0$. This follows directly from applying Eq. (19) to the permeability tensor in Eq. (1), which yields $\overline{\mu} \to \overline{\mu}^T$. In this case, the central region of the reciprocal dual platform forms a unidirectional waveguide that transports waves only from right to left, mirroring the behavior of the original system, which supports propagation solely from left to right in the relevant spectral range.

Let $\mathbf{f}_n(\mathbf{r};\alpha,\mathbf{B}_0)$ represent the family of eigenmodes of the original system associated with the eigenfrequencies $\omega_n(\alpha,\mathbf{B}_0)$. The explicit dependence on the dissipation parameter and external magnetic bias are indicated in the argument for clarity. In Appendix F, we show that the modes of the reciprocal dual system and those of the Hermitian-conjugate problem are related as follows:

$$\mathbf{f}_n(\mathbf{r};\alpha,-\mathbf{B}_0)=\boldsymbol{\sigma}\cdot\mathbf{f}_n^*(\mathbf{r};-\alpha,\mathbf{B}_0), \tag{20a}$$

$$\omega_n(\alpha,-\mathbf{B}_0)=\omega_n^*(-\alpha,\mathbf{B}_0)=\omega_n(\alpha,\mathbf{B}_0). \tag{20b}$$

It is therefore possible to construct the eigenfunctions of the reciprocal dual problem (with a flipped magnetic bias) from those of the Hermitian-conjugate problem (with negative dissipation). The physical significance of this result, in the present context, is as follows.

As seen in the previous subsection, the eigenfunctions of the Hermitian-conjugate problem are generally not square-integrable, since their divergent behavior near the relevant singular point is enhanced by the presence of gain. From Eq. (20), it is clear that the eigenfunctions of the reciprocal dual problem (with dissipation) share the same singularities as those of the Hermitian-conjugate problem (with gain). This leads to an important conclusion: dissipation alone does not regularize Maxwell's equations in the nonreciprocal cavity.

For the system in Fig. 1, supporting only left-to-right propagation under a magnetic bias oriented along +*z*, we found that point "R" becomes singular in the presence of negative dissipation. Equation (20) then implies that the same point becomes singular in the presence of positive dissipation when the magnetic bias is oriented along -*z*, corresponding to a waveguide that supports only right-to-left propagation.

By symmetry, the behavior of the eigenmodes of the reciprocal dual system near point "R" must mirror the behavior of the original system near point "L" in Fig. 1b). Therefore, we conclude that the eigenmodes of the original dissipative problem are not square-integrable near point "L" in the problematic spectral range, even though they remain square-integrable near point "R", as discussed earlier. Conversely, for the Hermitian-conjugate problem with gain, the situation is reversed: the eigenfunctions are square-integrable near point "L" but not near point "R".

This conclusion is confirmed by examining the asymptotic behavior of the eigenmodes of the reciprocal-dual system near point "R". Figure 4d) shows the singularity exponent of the fields near the wedge vertex for a magnetic field oriented along -*z*. One obtains exactly the same curves near point "L" for a magnetic bias directed along +*z*. As seen, in the presence of dissipation (green curves) the solution with $\mathrm{Re}\{\nu\}=0$ in the problematic spectral range moves into the region $\mathrm{Re}\{\nu\}<0$, confirming that dissipation enhances the divergent behavior in this case. In contrast, gain tames the divergence, shifting the singularity exponent into the region $\mathrm{Re}\{\nu\}>0$, as shown by the black dashed lines.

Without the explicit link between the eigenmodes of the reciprocal-dual and Hermitian-conjugate problems, it would not be obvious whether the branch of solutions with $\mathrm{Re}\{\nu\}=0$ near point "L" has any physical significance. One could argue that, since dissipation displaces the problematic branch into the unphysical region, it would not connect to actual eigenmodes, similar to other branches with $\mathrm{Re}\{\nu\}<0$. This was the point of view adopted in Ref. [15]. However, the connection between the reciprocal-dual and Hermitian-conjugate problems makes it clear that all the cavity eigenmodes in the problematic spectral range necessarily exhibit a square-integrable singularity near point "R" and a non-integrable singularity near point "L" for a magnetic bias along $+z$. The enhanced divergence at point "L" stems from the absence of a back-propagation path and the consequent inability to form standing waves. Without such a path, it is impossible to construct eigenmodes that remain finite at point "L", since this point effectively acts as the source of the radiation propagating toward point "R".

Consistent with this picture, one can verify using the same arguments as in the previous section that the heat current also exhibits a line-type singularity at point "L", with $\mathbf{S}_\omega(\mathbf{r}) \sim \frac{1}{\rho}$ in the relevant spectral range. Clearly, point "L" behaves as the source of thermal energy that is absorbed at point "R", further underscoring the persistence of the thermodynamic paradox as a manifestation of the ill-posedness of the eigenmodes in the nonreciprocal cavity.

For conventional localized excitations, the non-integrable singularity is largely inconsequential: the unidirectional nature of the waveguide carries radiation away from point "L". In contrast, for fluctuation induced fields, where noise currents are distributed throughout the material, the singularity at point "L" becomes a dominant feature of the equilibrium response.

It is worth noting that there are many examples in the literature where reciprocal plasmonic platforms with sharp boundaries display non-integrable singular fields in the lossless limit [34-38]. However, for reciprocal systems, any finite dissipation always regularizes the stored energy, rendering the fields square-integrable regardless of the wedge geometry (see Appendix A in Ref. [34]). In contrast, our analysis reveals that nonreciprocal systems are far more subtle: singular behaviors cannot be regularized by dissipation alone.

## IV. Resolution of the Paradox

Next, we show that nonlocal effects provide a natural resolution to the thermodynamic paradox. Nonlocality has been extensively discussed in the photonics literature, most often in connection with the regularization of plasmonic responses in metals and semiconductors [39-43]. A widely used framework in this context is the hydrodynamic description of a plasma, in which nonlocality arises from repulsive electron–electron interactions, introducing a diffusion term into the electron transport equation. Numerous studies have demonstrated how this diffusive behavior suppresses unphysical singularities in plasmonic nanostructures with sharp wedges [41-43], even in the limit of vanishing dissipation. This property stems from the fact that diffusive effects are driven by a force proportional to the gradient of the charge density in the plasma, making them especially effective in regions where the fields tend to concentrate. While nonlocality smooths the fields and removes divergences, it can still give rise to "hotspots" with strong field localization, though the field amplitudes remain finite in such regions [41-43]. Another form of nonlocality discussed in metallic systems relates to the so-called "odd viscosity" studied by Vitelli and co-workers [44].

In contrast to metallic platforms, nonlocal effects in a magnetically biased ferrite have received relatively little attention. In the next subsection, we introduce a minimal model in which nonlocality does not emerge from a specific microscopic mechanism, such as electron–electron repulsion in a plasma, but rather from a universal feature of macroscopic electromagnetism, specifically it is an effective medium description of a granular material, typically possessing an intrinsic periodicity at the atomic scale.

### *A. Intrinsic nonlocality from spatial averaging*

We begin by briefly reviewing the response theory of a magnetized medium composed of magnetic dipoles. The dynamics of a magnetic dipole moment (spin) under an external magnetic field is governed by the Landau–Lifshitz equation, $\frac{d\mathbf{m}}{dt} = -\frac{e}{m}\mathbf{m} \times \mu_0 \mathbf{H}$ (for simplicity, we take the *g*-factor to be exactly 2). Here, $-e$ is the electron charge and *m* the electron (effective) mass.

In the presence of a static magnetic bias ($\mathbf{B}_0 = \mu_0 \mathbf{H}_0$), the response to a weak time-varying magnetic field is obtained by linearizing the Landau–Lifshitz equation, ($\mathbf{m} \to \mathbf{m}_0 + \mathbf{m}$, $\mathbf{H} \to \mathbf{H}_0 + \mathbf{H}$), which yields:

$$\frac{d\mathbf{m}}{dt} = -\frac{e}{m}\mu_0 \mathbf{m}_0 \times \mathbf{H} - \mathbf{m} \times \frac{e}{m}\mathbf{B}_0 . \tag{21}$$

The above formula controls the (linearized) motion of a given magnetic dipole. The macroscopic magnetization $\mathbf{M}$ is obtained by spatially averaging the sum of the contributions from all magnetic dipoles $\sum_i \mathbf{m}_i \delta(\mathbf{r} - \mathbf{r}_i)$. Here, $\mathbf{m}_i$ represents the magnetic dipole of a generic particle positioned at $\mathbf{r}_i$.

In the local approximation, the magnetization is taken as $\mathbf{M}(\mathbf{r}) \approx \mathbf{m}_i / V_{\text{cell}}$ for an observation point $\mathbf{r}$ in the vicinity of the *i*th particle. Here, $V_{\text{cell}}$ is the volume of the unit cell. For simplicity, we assume the dipoles are arranged in a periodic lattice. This leads to the well-known master equation [1]:

$$\frac{d\mathbf{M}}{dt} = -\boldsymbol{\omega}_{\mathrm{m}} \times \mathbf{H} - \mathbf{M} \times \boldsymbol{\omega}_0 , \qquad (22)$$

where for convenience we denote $\boldsymbol{\omega}_0 = \frac{e}{m}\mathbf{B}_0$ as the cyclotron (Larmor) frequency vector, and $\boldsymbol{\omega}_{\mathrm{m}} = \frac{e}{m}\frac{\mu_0 \mathbf{m}_0}{V_{\mathrm{cell}}}$ as the cyclotron frequency vector for the saturation magnetization ($\mathbf{M}_0 = \frac{\mathbf{m}_0}{V_{\mathrm{cell}}}$). When the magnetic bias is aligned with the *z*-direction ($\mathbf{B}_0 \sim \hat{\mathbf{z}}$), one can write $\boldsymbol{\omega}_0 = \omega_0 \hat{\mathbf{z}}$ and $\boldsymbol{\omega}_{\mathrm{m}} = \omega_{\mathrm{m}} \hat{\mathbf{z}}$, with $\omega_0, \omega_{\mathrm{m}}$ sharing the same sign (positive for a magnetic field along +*z*). The fields in the material can be characterized by combining Eq. (22) with the macroscopic Maxwell's equations:

$$-\nabla \times \mathbf{E} = \mu_0 \partial_t \mathbf{H} + \mathbf{J}_{\mathrm{s}} , \qquad \nabla \times \mathbf{H} = \varepsilon_0 \varepsilon \, \partial_t \mathbf{E} , \qquad (23)$$

with $\mathbf{J}_{\mathrm{s}} = \mu_0 \partial_t \mathbf{M}$ the equivalent magnetic current and $\partial_t \equiv \partial / \partial t$. For time-harmonic time-variations of the type $e^{-i\omega t}$, such an approach is formally equivalent to introducing a magnetic permeability defined as in Eq. (1). Dissipation has been neglected in Eq. (22), so the equivalence strictly holds in the lossless limit [1].

While the above approach is accurate in the long-wavelength limit, it breaks down for field distributions with significant variation across a single unit cell. Macroscopic electromagnetism is, by construction, a theory of spatially averaged fields, in which microscopic fluctuations on the scale of the unit cell are inherently smoothed out by the homogenization process [45-48]. The purely local model of Eq. (1) ignores this essential feature: it imposes no constraint on the degree of spatial localization in the material response, thereby allowing unphysical singularities to emerge, precisely the type of behavior identified in Sect. III. These extreme localizations can appear even when the excitation fields are smooth and delocalized, in direct violation of the foundational principles of homogenization theory [45-48].

As described by Russakoff [45, 46], a more rigorous way to determine the material's macroscopic response is to define it through a convolution with a test function $f(\mathbf{r})$ localized around the unit cell (assumed to be centered at the origin). Specifically, for a generic field $\mathbf{A}(\mathbf{r})$, the corresponding macroscopic field is given by

$$\langle \mathbf{A}(\mathbf{r},t) \rangle = \iiint \mathbf{A}(\mathbf{r}-\mathbf{r}',t) f(\mathbf{r}') d^3\mathbf{r}' . \quad (24)$$

Applying this procedure to the ferrite, the macroscopic magnetization should be taken as $\mathbf{M} = \left\langle \sum_i \mathbf{m}_i \delta(\mathbf{r}-\mathbf{r}_i) \right\rangle \approx \langle \mathbf{M}_{\text{loc}} \rangle$, obtained by averaging the magnetization $\mathbf{M}_{\text{loc}}$ predicted by the local formalism of Eq. (22). As the averaging operation is linear, $\mathbf{M}$ is now governed by the modified master equation:

$$\frac{d\mathbf{M}}{dt} = -\boldsymbol{\omega}_{\text{m}} \times \langle \mathbf{H}_{\text{loc}} \rangle - \mathbf{M} \times \boldsymbol{\omega}_0 , \quad (25)$$

where $\langle \mathbf{H}_{\text{loc}} \rangle$ is a smoothed version of the local field acting on the magnetic dipoles. Evidently, in the Fourier domain ($\mathbf{r} \leftrightarrow \mathbf{k}$) this averaging can be expressed as $\langle \mathbf{H}_{\text{loc}} \rangle(\mathbf{k}) = \mathbf{H}(\mathbf{k}) F(\mathbf{k})$ where $F(\mathbf{k})$ represents the Fourier transform of the test function *f*, acting as a spatial low-pass filter. For concreteness, we choose $F(\mathbf{k}) = 1/\left(1 + k^2 / k_{\max}^2\right)$, where $k_{\max}$ is a cut-off parameter controlled by the largest dimension $a_0$ of the unit cell ($k_{\max} \sim \pi / a_0$). This choice is convenient because the inverse Fourier transform yields an explicit relation between $\langle \mathbf{H}_{\text{loc}} \rangle$ and $\mathbf{H}$ in the spatial domain:

$$\left[-\frac{1}{k_{\max}^2}\nabla^2+1\right]\langle \mathbf{H}_{\rm loc}\rangle = \mathbf{H} . \tag{26}$$

Therefore, the intrinsic granularity of the medium implies an unavoidable nonlocal response for wavelengths comparable to the lattice constant. This effect can be captured by replacing the local model of Eq. (22) with the modified spatially averaged formulation in Eqs. (25)-(26). To ensure uniqueness of the solution of Eq. (26), we impose that $\langle \mathbf{H}_{\rm loc}\rangle$ vanishes at the ferrite boundary. According to Eq. (25), this requirement is equivalent to enforcing that the magnetization vector $\mathbf{M}$ itself vanishes at the ferrite boundary. It can be verified that the boundary condition $\mathbf{M}=0$ is consistent with energy conservation in the time-harmonic regime. The need for additional boundary conditions is a general feature of nonlocal systems. For instance, in the hydrodynamic model of an electron gas, the normal component of the electric current must vanish at the boundary [39-43].

It is worth emphasizing that the proposed wave vector cut-off regularization is quite general and can be readily extended to a variety of reciprocal and nonreciprocal material platforms, including magnetized plasmas, magnetized semiconductors, and related systems. In such cases, the nonlocal effects stemming from the intrinsic granularity of the medium may compete or interplay with other sources of nonlocality, such as the diffusion-driven effects discussed earlier. For a magnetized ferrite, $k_{\max}$ may be linked to the exchange length $l_{\rm ex}$, which is the physical length scale that limits the continuous-medium approximation ($k_{\max} \sim 1/l_{\rm ex}$) [49]. It corresponds to the shortest meaningful wavelength for spin excitations (magnons) before continuum micromagnetics breaks down, and is typically on the order of 10–20 nm.

Combining Eqs. (25)-(26), and switching to the Fourier domain ($(t,\mathbf{r}) \leftrightarrow (\omega,\mathbf{k})$) one readily finds that the equivalent magnetic permeability with the wave vector cut-off $k_{\max}$ is determined by:

$$\overline{\mu}_{\rm nl}(\omega,\mathbf{k}) = \mu_0 \mathbf{1} + \frac{1}{1+k^2/k_{\max}^2}\left[\overline{\mu}(\omega) - \mu_0 \mathbf{1}\right], \tag{27}$$

where $\overline{\mu}(\omega)$ represents the local response given by Eq. (1) and the term $F(\mathbf{k}) = 1/\left(1+k^2/k_{\max}^2\right)$ is the low-pass filter introduced by the averaging procedure. For short wavelengths (large *k*), the material response is effectively suppressed by $F(\mathbf{k})$, which enforces the intrinsic resolution limit set by the unit-cell scale. This form of nonlocality has been widely employed as a means of regularizing the topology of the material response in electromagnetic continua [50].

In Appendix E, we demonstrate that in the presence of nonlocality the singularity exponents $\nu$ governing the asymptotic field behavior become integer values, independent of the ferrite dispersion or material dissipation. Thus nonlocality completely suppresses the electromagnetic singularities in our system. As further detailed in Appendix E, this behavior, markedly different from the local model, originates from the increased regularity of the magnetic field $\mathbf{H}$ across the ferrite–air interface, which becomes a continuous function (normal component included) due the vanishing of the magnetization vector $\mathbf{M}$ at the ferrite boundary. In this way, nonlocality offers a straightforward physical resolution of the thermodynamic paradox by eliminating field localization on spatial scales smaller than $1/k_{\max}$. The singularity suppression in our model is entirely analogous to that observed in plasmonic nanostructures under the hydrodynamic model [41-43], where diffusion effects enforce continuity of the electric field at a metal–dielectric interface.

From a different perspective, the heat-current singularities are associated with large-$k$ spatial harmonics. In our model, the wave-vector cut-off suppresses the material response at high $k$, causing the system to behave near the wedge vertex as a reciprocal medium, for which the Poynting-vector expectation vanishes. Consequently, the cut-off regularizes the thermodynamic response by forcing the heat current to approach zero near the vertex, thereby removing the unphysical singularity. This argument also makes it clear that the proposed regularization is not sensitive to the specific form of the kernel $F(\mathbf{k})$, as long as it ensures that the material response becomes reciprocal at large $k$.

To illustrate further the impact of the nonlocality, next we focus on the most challenging scenario: vanishingly small material dissipation, where the singularity near point "R" in Fig. 1 is most pronounced in the local case. As we show below, nonlocality prevents field concentration by enabling a return channel for the thermally generated radiation propagating from left to right, thereby rendering the waveguide effectively bidirectional and eliminating the spectral window responsible for the pathological unidirectional behavior.

Figure 2 shows the dispersion of the extra guided wave produced by nonlocal effects, which propagates in the negative $x$-direction (black lines, for different values of $k_{\max}$). Its group velocity decreases steadily as the cut-off parameter $k_{\max}$ is increased, tending to zero as $k_{\max} \to \infty$, with the guided wavenumber diverging in the same limit. In this regime, the mode becomes increasingly confined to the bottom wall of the waveguide ($y$=0), mirroring the trends reported in Refs. [15, 32] for related nonlocal systems.

Physically, as the nonlocality weakens, the guided wave in the return channel becomes slower and more tightly bound to the boundary, but crucially, it still exists, ensuring that no trapped energy builds up at point “R” and that the fields remain finite.

The dispersion curves in Fig. 2 were obtained by numerically solving Eqs. (23), (25) and (26) in the frequency domain, with PEC boundary conditions on the metallic walls ($E_z = 0$) and $\mathbf{M} = 0$ at the ferrite interface. These additional boundary conditions ($M_x = M_y = 0$) are consistent with the fact that nonlocality induces 2 additional TE waves in the bulk medium. The analytical procedure used to find the edge mode follows closely the methods described in Refs. [51, 52] and is not repeated here. Because the additional guided wave is strongly confined to the bottom plate, the ferrite thickness was taken effectively infinite in the calculations. We numerically verified that the edge state still emerges and remains gapless for a Neumann-type additional boundary condition, $\frac{\partial M_x}{\partial y} = \frac{\partial M_y}{\partial y} = 0$ (not shown).

### *B. Topological protection and the paradox*

The elimination of the thermodynamic paradox through nonlocality is deeply connected to the emergence of topological protection. As shown in previous work [32, 50], once nonlocal effects are included the magnetized ferrite develops a well-defined topological gap in the problematic spectral range, characterized by a nonzero gap Chern number $\mathcal{C}_{\text{gap}} = \text{sgn}(\omega_0)$. The gap Chern number is robust to the effect of dissipation [53]. The additional guided mode that emerges (black lines in Fig. 2) is precisely the gapless topological edge state mandated by bulk–boundary correspondence at the interface between the PEC and the magnetized ferrite. This edge state provides the missing return channel, which renders the waveguide effectively bidirectional and thereby removes the pathological unidirectional propagation responsible for the paradox. We note in passing that the regularization of material topology has also been discussed in the context of the hydrodynamic plasma model [51, 54, 55, 56].

In the local model, where spatial-dispersion is neglected, the material topology is ill-defined [50, 51, 52], the edge state is suppressed, and the paradox persists. Consistent with the findings of this work, it has been shown that dissipation alone cannot generate a well-defined topology, even when nonreciprocal elements are arranged in a lattice [57, 58]. Thus, real-space non-integrable singularities of the fields can be viewed as the direct physical footprint of topological ill-posedness, both in Hermitian and dissipative systems. By contrast, once nonlocality is restored, the topology becomes well defined and the edge state necessarily appears, ensuring the removal of the thermodynamic inconsistency.

The absence of real-space singularities in Hermitian systems is essentially equivalent to ensuring that the bulk edge correspondence must hold true. In fact, the bulk-edge correspondence implies that at a junction of different non-dissipative topological materials the number of edge states propagating towards a junction point must be exactly identical to the number of edge states propagating away from it [14, 15, 32, 39]. If that were not true, it would be possible to consider excitations of the system (or of its reciprocal dual) for which there would be a wave propagating towards the junction but no wave emerging from it, implying the existence of a real-space singularity at the junction point [14, 39].

In non-Hermitian systems, the constraints imposed by the absence of singularities remain subtle yet nontrivial. To illustrate this, consider the geometry sketched in Fig. 5: a lossy magnetized ferrite with a wave-vector cutoff enclosed within a cavity whose boundary alternates between perfect electric (PEC) and perfect magnetic (PMC) conductor walls. Suppose that this cavity supports an electromagnetic mode $\mathbf{f}_n(\mathbf{r};\alpha)$ with complex frequency $\omega_n = \omega_n' + i\omega_n''$ lying in the lower-half frequency plane in a spectral gap of the bulk region. The contribution of this mode to the spectral density of the heat current is determined by Eq. (13), through the mixed Poynting vector $\mathbf{S}_n(\mathbf{r};\alpha)$ constructed from the mode and its Hermitian-conjugate partner $\mathbf{f}_n(\mathbf{r};-\alpha)$.

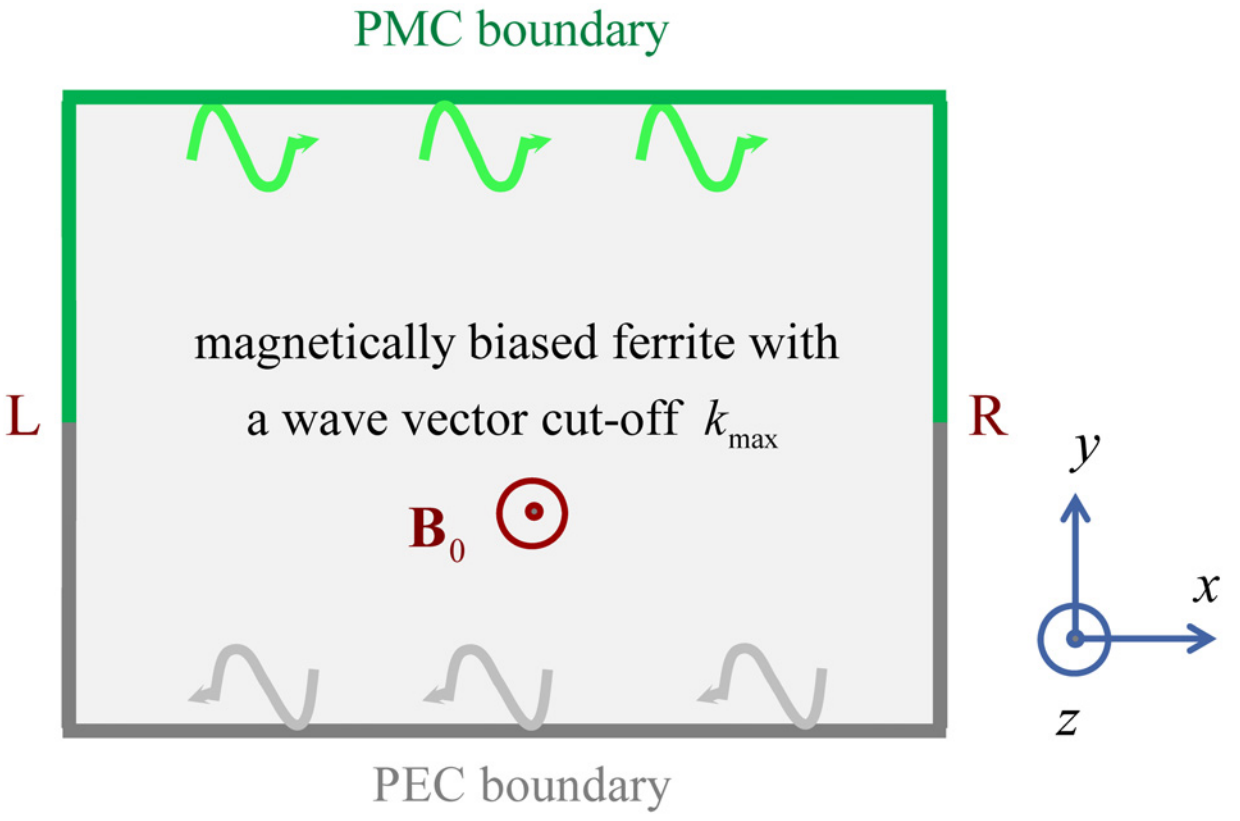

**Fig. 5** Two-dimensional cavity formed by a magnetically biased ferrite with a wave-vector cut-off, bounded below by a PEC wall and above by a PMC wall. The PEC–ferrite interface supports a damped edge state, generating cavity modes with a nontrivial mixed Poynting vector $\mathbf{S}_n(\mathbf{r};\alpha)$. To ensure closed lines and avoid thermal sinks or sources, the ferrite–PMC interface must also support edge-states propagating in the same direction.

The key point is that nonlocality guarantees that $\mathbf{S}_n(\mathbf{r};\alpha)$ is free of singularities. Consequently, its vector lines must be closed, as neither sinks nor sources are admissible and $\nabla\cdot\mathbf{S}_n = 0$ [Eq. (D3)]. Because $\omega_n$ lies, by hypothesis, within a spectral gap of the bulk medium, the corresponding fields must be exponentially localized at the cavity walls. This constraint applies equally to the dissipative system and to its Hermitian-conjugate counterpart with gain. Consequently, the streamlines of the mixed Poynting vector $\mathbf{S}_n(\mathbf{r};\alpha)$ are likewise exponentially confined to the cavity boundaries.

This observation has direct implications. As discussed previously, the ferrite–PEC interface supports a single topological edge-state branch in the presence of nonlocality. In the dissipative case, this edge state becomes damped, yet it still produces cavity modes confined to the PEC wall (albeit with a damping constant that usually differs from the flat-interface case). We argue that the ferrite–PMC interface must also support an edge-state branch propagating in the same clockwise direction. If this were not the case, it would be impossible to close the streamlines of the mixed Poynting vector associated with the cavity mode, leading to a contradiction. Consistency therefore requires that the ferrite–PMC interface host at least one clockwise-propagating edge state branch, ensuring that thermal energy circulates around the cavity without generating sinks or sources.

Whether this reasoning can be elevated to a general principle, namely that the net number of damped unidirectional edge states across all interfaces must always balance as in the Hermitian case, remains unclear. It is well established that bulk–edge correspondence can break down in non-Hermitian systems due to the so-called non-Hermitian skin effect [59-61]. However, such breakdowns typically occur in systems combining gain and loss. In the presence of gain, there is no fundamental reason to exclude field singularities or thermal sinks and sources, as energy injection may sustain them. By contrast, the fully passive case considered here is qualitatively different: singular field behavior is forbidden, as it would be incompatible with the conditions required for thermodynamic equilibrium. Thus, while the generalization remains unresolved, our analysis demonstrates that in dissipative nonreciprocal systems the exclusion of real-space singularities already enforces a nontrivial bulk–edge constraint: in a closed cavity, the mixed Poynting vector associated with any eigenmode must form closed loops, and this in turn requires a consistent set of edge states across all boundary types.

## V. Summary

We have revisited the long-standing thermodynamic paradox associated with unidirectional modes in nonreciprocal waveguides and cavities. Through a rigorous modal analysis, we showed that dissipation alone does not regularize the singular behavior of the eigenmodes, in sharp contrast with reciprocal systems where any finite loss renders the fields square-integrable near sharp wedges. In nonreciprocal cavities, singularities persist at specific transition points, reflecting the absence of back-propagating channels and preventing the formation of standing waves. These pathologies manifest directly in the fluctuation–dissipation response, giving rise to unphysical sources and sinks of thermal energy.

We demonstrated that resolving the paradox requires the inclusion of nonlocal effects, which naturally suppress the material response at short wavelengths. In particular, we showed that nonlocality rooted in spatial averaging and macroscopic homogenization renders the system effectively bidirectional in the problematic spectral range. This mechanism enables the formation of a return channel associated with a topological edge state, thereby eliminating the thermodynamic inconsistency. In this sense, nonlocality and topology are deeply intertwined: the removal of real-space singularities correlates with the restoration of a well-defined topology. Furthermore, by analyzing the mixed Poynting vector constructed from eigenmodes and their Hermitian-conjugate partners, we argued that the absence of singularities in the dissipative case imposes strong constraints on the existence and circulation of edge modes in closed cavities. While a full generalization of bulk–edge correspondence to passive non-Hermitian systems remains open, our results suggest that thermodynamic consistency enforces nontrivial restrictions that closely parallel the Hermitian case.

A relevant experimental validation of the present theory would be to confirm that, in an undriven cavity, there is no net flow of electromagnetic energy between points "L" and "R" (i.e., no temperature gradient develops across the two points as a result of the nonreciprocal bias), thereby demonstrating that nonlocality suppresses non-integrable singularities and ensures thermodynamic consistency.

In summary, our findings establish a direct link between thermodynamic consistency, nonlocal regularization, and topological protection in nonreciprocal photonic systems, showing that ill-posed topologies and real-space singularities are two sides of the same coin.

**Acknowledgements:** This work is supported in part by the Institution of Engineering and Technology (IET), by the Simons Foundation (award SFI-MPS-EWP-00008530-10), and by national funds through FCT – Fundação para a Ciência e a Tecnologia, I.P., and, when eligible, co-funded by EU funds under project/support UID/50008/2025 – Instituto de Telecomunicações, with DOI identifier https://doi.org/10.54499/UID/50008/2025.

## Appendix A: Electrodynamics of non-Hermitian dispersive systems

In this appendix, we briefly review the theory presented in Ref. [53, Appendix C], which demonstrates that the electrodynamics of generic dispersive dissipative materials, characterized by a meromorphic material matrix $\mathbf{M}$, can always be reformulated as a standard time-evolution problem. Related results for conservative (Hermitian) systems can be found in Refs. [50, 62]. In addition, we formally introduce the Hermitian-conjugate problem and establish its connection to the response of the original system.

***Extended problem***

The frequency domain Maxwell's equations in a dispersive material platform can be written in a compact form as:

$$\hat{\mathcal{D}}\cdot\mathbf{f}=\omega\mathbf{M}(\mathbf{r},\omega)\cdot\mathbf{f}+i\mathbf{j}, \tag{A1}$$

where $\mathbf{f}=\begin{pmatrix}\mathbf{E} & \mathbf{H}\end{pmatrix}^T$ is a six-component vector containing the electromagnetic fields, $\mathbf{j}=\begin{pmatrix}\mathbf{j}_e & \mathbf{j}_m\end{pmatrix}^T$ is a six-component vector of electric and magnetic currents, and $\hat{\mathcal{D}}$ is the differential operator given by Eq. (9).

The material matrix $\mathbf{M}$ relates the frequency-domain macroscopic fields via the constitutive relations as in Eq. (6). We assume that it admits a partial-fraction decomposition of the form:

$$\mathbf{M}(\mathbf{r},\omega)=\mathbf{M}_\infty+\sum_\alpha\frac{\mathbf{R}_\alpha}{\omega-\omega_{p,\alpha}}. \tag{A2}$$

Here, $\mathbf{M}_\infty=\lim_{\omega\to\infty}\mathbf{M}(\omega)$ gives the asymptotic high-frequency response of the material (typically coincident with the response of the vacuum, i.e., $\mathbf{M}_\infty=\begin{pmatrix}\varepsilon_0\mathbf{1}_{3\times3} & 0\\ 0 & \mu_0\mathbf{1}_{3\times3}\end{pmatrix}$). Furthermore, $\omega_{p,\alpha}$ are the (complex-valued) poles of $\mathbf{M}$ and $\mathbf{R}_\alpha=\lim_{\omega\to\omega_{p,\alpha}}\mathbf{M}(\mathbf{r},\omega)(\omega-\omega_{p,\alpha})$ is the residue for the $\omega_{p,\alpha}$ pole. For passive materials, these poles lie in the lower half of the complex frequency plane.

Following Ref. [53], it is useful to introduce the auxiliary fields

$$\mathbf{Q}^{(\alpha)}(\mathbf{r},\omega)=\frac{\left(s_\alpha\omega_{p,\alpha}\right)^{1/2}}{\left(\omega-\omega_{p,\alpha}\right)}\mathbf{A}_\alpha\cdot\mathbf{f}(\mathbf{r},\omega), \tag{A3}$$

with $s_\alpha=\operatorname{sgn}\left(\operatorname{Re}\left\{\omega_{p,\alpha}\right\}\right)$ and $\mathbf{A}_\alpha=\left[-s_\alpha\mathbf{R}_\alpha\right]^{1/2}$.

Then, the time-dynamics implicitly defined by Eq. (A1) can be explicitly formulated in terms of a generalized state vector $\mathbf{Q}=\begin{pmatrix}\mathbf{f} & \mathbf{Q}^{(1)} & \ldots & \mathbf{Q}^{(\alpha)} & \ldots\end{pmatrix}^T$ as follows [53]:

$$\hat{L}\cdot\mathbf{Q}(\mathbf{r},t)=i\frac{\partial}{\partial t}\mathbf{M}_{\mathrm{g}}\cdot\mathbf{Q}(\mathbf{r},t)+i\mathbf{j}_{\mathrm{g}}(\mathbf{r},t), \tag{A4}$$

with

$$\hat{L}=\begin{pmatrix}\hat{\mathcal{D}}+\sum\limits_\alpha s_\alpha\mathbf{A}_\alpha^2 & \left(s_1\omega_{p,1}\right)^{1/2}\mathbf{A}_1 & \left(s_2\omega_{p,2}\right)^{1/2}\mathbf{A}_2 & \ldots\\ \left(s_1\omega_{p,1}\right)^{1/2}\mathbf{A}_1 & \omega_{p,1}\mathbf{1} & \mathbf{0} & \ldots\\ \left(s_2\omega_{p,2}\right)^{1/2}\mathbf{A}_2 & \mathbf{0} & \omega_{p,2}\mathbf{1} & \ldots\\ \ldots & \ldots & \ldots & \ldots\end{pmatrix},\qquad \mathbf{M}_{\mathrm{g}}=\begin{pmatrix}\mathbf{M}_\infty & \mathbf{0} & \mathbf{0} & \ldots\\ \mathbf{0} & \mathbf{1} & \mathbf{0} & \ldots\\ \mathbf{0} & \mathbf{0} & \mathbf{1} & \ldots\\ \ldots & \ldots & \ldots & \ldots\end{pmatrix}. \tag{A5}$$

In the above, $\mathbf{j}_g=\begin{pmatrix}\mathbf{j} & \mathbf{0} & \mathbf{0} & \ldots\end{pmatrix}^T$ is a generalized current.

For conservative materials ($\mathbf{M}=\mathbf{M}^\dagger$ for $\omega$ real-valued), the poles are real-valued and $\mathbf{A}_\alpha$ are non-negative Hermitian matrices [28, 50]. In such a case, the operator $\hat{L}$ is Hermitian with respect to the canonical inner product: $\hat{L}=\hat{L}^\dagger$. In contrast, for dissipative systems $\hat{L}\neq\hat{L}^\dagger\equiv\hat{L}^{\mathrm{c}}$.

***Hermitian conjugated problem***

It is useful to introduce a Hermitian conjugate time-domain problem defined by

$$\hat{L}^{\mathrm{c}}\cdot\mathbf{Q}^{\mathrm{c}}=i\frac{\partial}{\partial t}\mathbf{M}_{\mathrm{g}}\cdot\mathbf{Q}^{\mathrm{c}}+i\mathbf{j}_{\mathrm{g}}^{\mathrm{c}}. \tag{A6}$$

It is straightforward to verify that this equation describes a material with the following dispersive response:

$$\mathbf{M}^{\mathrm{c}}(\omega) = \mathbf{M}_{\infty} + \sum_{\alpha} \frac{\mathbf{R}_{\alpha}^{\dagger}}{\omega - \omega_{p,\alpha}^{*}} = \left[\mathbf{M}\left(\omega^{*}\right)\right]^{\dagger}. \quad \text{(A7)}$$

Specifically, we decompose $\mathbf{Q}^{\mathrm{c}}$ as $\mathbf{Q}^{\mathrm{c}} = \begin{pmatrix}\mathbf{f}^{\mathrm{c}} & \mathbf{Q}^{\mathrm{c},(1)} & ...\end{pmatrix}^{T}$ and take an excitation of the type $\mathbf{j}_{\mathrm{g}}^{\mathrm{c}} = \begin{pmatrix}\mathbf{j}^{\mathrm{c}} & \mathbf{0} & \mathbf{0} & ...\end{pmatrix}^{T}$, then the electromagnetic component $\mathbf{f}^{\mathrm{c}}$ of the state vector satisfies

$$\hat{\mathcal{D}} \cdot \mathbf{f}^{\mathrm{c}} = \omega \mathbf{M}^{\mathrm{c}}(\mathbf{r}, \omega) \cdot \mathbf{f}^{\mathrm{c}} + i\mathbf{j}^{\mathrm{c}}, \quad \text{(A8)}$$

in the frequency domain. Thus, the Hermitian conjugate problem describes the electrodynamics of a material whose response is the Hermitian conjugate of the original system, typically representing a gain medium.

For future reference, we consider the spectral problem associated with Eq. (A4):

$$\hat{H}_{\mathrm{g}} \mathbf{Q}_{n} = \omega_{n} \mathbf{Q}_{n}, \qquad \hat{H}_{\mathrm{g}} = \mathbf{M}_{\mathrm{g}}^{-1} \cdot \hat{L}. \quad \text{(A9a)}$$

Here, $\mathbf{Q}_n$, $\omega_n$ are the eigenfunctions and eigenvalues of $\hat{H}_{\mathrm{g}}$ ($n$=1, 2, …). Similarly, the spectral problem for the Hermitian conjugate system [Eq. (A6)] is:

$$\hat{H}_{\mathrm{g}}^{\mathrm{c}} \mathbf{Q}_{n}^{\mathrm{c}} = \omega_{\mathrm{c},n} \mathbf{Q}_{n}^{\mathrm{c}}, \qquad \hat{H}_{\mathrm{g}}^{\mathrm{c}} = \mathbf{M}_{\mathrm{g}}^{-1} \cdot \hat{L}^{\mathrm{c}}. \quad \text{(A9b)}$$

It is convenient to introduce the weighted inner product defined by:

$$\left\langle \mathbf{Q}_{B} \mid \mathbf{Q}_{A} \right\rangle \equiv \int_{V} \frac{1}{2} \mathbf{Q}_{B}^{*} \cdot \mathbf{M}_{\mathrm{g}} \cdot \mathbf{Q}_{A} d^{3}\mathbf{r}. \quad \text{(A10)}$$

It is implicit that $\mathbf{M}_{\mathrm{g}}$ is real-valued, symmetric and positive definite. Then, it is straightforward to verify that $\left\langle \mathbf{Q}_B \,|\, \hat{H}_{\mathrm{g}} \mathbf{Q}_A \right\rangle = \left\langle \hat{H}_{\mathrm{g}}^{\mathrm{c}} \mathbf{Q}_B \,|\, \mathbf{Q}_A \right\rangle$, i.e., the operators $\hat{H}_{\mathrm{g}}, \hat{H}_{\mathrm{g}}^{\mathrm{c}}$ are Hermitian conjugate with respect to the weighted inner product.

## Appendix B: Modal expansions in dispersive non-Hermitian systems

In this appendix, we develop bi-orthogonal modal expansions of the electromagnetic fields in generic dispersive non-Hermitian platforms.

***Bi-orthogonal expansion***

Let $\mathbf{Q}_n$, $\omega_n$ ($\mathbf{Q}_n^{\mathrm{c}}$, $\omega_{\mathrm{c},n}$) be the eigenfunctions and eigenvalues of the operator $\hat{H}_{\mathrm{g}}$ ($\hat{H}_{\mathrm{g}}^{\mathrm{c}}$), respectively, defined as in Eq. (A9) of Appendix A. Noting that $\hat{H}_{\mathrm{g}}$ and $\hat{H}_{\mathrm{g}}^{\mathrm{c}}$ are Hermitian conjugate with respect to the weighted inner product (A10), it follows that $\det\left(\hat{H}_{\mathrm{g}} - \omega \mathbf{1}\right) = \det\left(\hat{H}_{\mathrm{g}}^{\mathrm{c}} - \omega^* \mathbf{1}\right)^*$. Thus, the eigenvalues of $\hat{H}_{\mathrm{g}}$ and $\hat{H}_{\mathrm{g}}^{\mathrm{c}}$ are linked by complex conjugation. In particular, the eigenfunctions can be ordered in such way that:

$$\omega_n = \omega_{\mathrm{c},n}^* . \tag{B1}$$

As $\left\langle \mathbf{Q}_m^{\mathrm{c}} \,|\, \hat{H}_{\mathrm{g}} \mathbf{Q}_n \right\rangle = \left\langle \hat{H}_{\mathrm{g}}^{\mathrm{c}} \mathbf{Q}_m^{\mathrm{c}} \,|\, \mathbf{Q}_n \right\rangle$, we have that $\left(\omega_{\mathrm{c},m}^* - \omega_n\right)\left\langle \mathbf{Q}_m^{\mathrm{c}} \,|\, \mathbf{Q}_n \right\rangle = 0$. Hence, the eigenmodes satisfy:

$$\left\langle \mathbf{Q}_m^{\mathrm{c}} \,|\, \mathbf{Q}_n \right\rangle = 0 , \qquad \text{when} \quad \omega_{\mathrm{c},m}^* \neq \omega_n . \tag{B2}$$

When the two sets of eigenvectors are complete (which is typically the case when $\hat{H}_{\mathrm{g}}$ is a sufficiently weak perturbation of a Hermitian system), the eigenmodes can be chosen to satisfy the generalized orthogonality relations:

$$\left\langle \mathbf{Q}_m^{\mathrm{c}} \mid \mathbf{Q}_n \right\rangle = \delta_{m,n} . \tag{B3}$$

Thus, $\mathbf{Q}_n$ and $\mathbf{Q}_n^{\mathrm{c}}$ form a bi-orthogonal system [26, 27]. In particular, a generic state vector can be expanded as:

$$\mathbf{Q}(\mathbf{r}) = \sum_n \mathbf{Q}_n(\mathbf{r}) c_n \qquad \text{with} \quad c_n = \left\langle \mathbf{Q}_n^{\mathrm{c}} \mid \mathbf{Q} \right\rangle . \tag{B4}$$

***Modal Expansion of the Electromagnetic Fields***

Next, we show that a generic solution $\mathbf{f}$ of the source-driven time-harmonic Maxwell equations in a non-Hermitian system [Eq. (A1)] admits the following modal expansion:

$$\mathbf{f}(\mathbf{r}) = \sum_n \mathbf{f}_n(\mathbf{r}) c_n , \qquad \text{with} \quad c_n = \frac{1}{2} \int_V d^3\mathbf{r} \frac{\mathbf{f}_n^{\mathrm{c}*} \cdot i\mathbf{j}}{\omega_n - \omega} . \tag{B5}$$

Here, $\mathbf{f}_n(\mathbf{r})$ ($\mathbf{f}_n^{\mathrm{c}}(\mathbf{r})$) are the eigenmodes of the system governed by the material matrix $\mathbf{M}$ ($\mathbf{M}^{\mathrm{c}}$). In particular, one has

$$\hat{\mathcal{D}} \cdot \mathbf{f}_n = \omega_n \mathbf{M}(\mathbf{r}, \omega_n) \cdot \mathbf{f}_n , \quad \hat{\mathcal{D}} \cdot \mathbf{f}_n^{\mathrm{c}} = \omega_{\mathrm{c},n} \mathbf{M}^{\mathrm{c}}\left(\mathbf{r}, \omega_{\mathrm{c},n}\right) \cdot \mathbf{f}_n^{\mathrm{c}} , \tag{B6}$$

which correspond to Eqs. (A1) and (A8), with trivial excitations.

To demonstrate Eq. (B5), it is convenient to consider the solution $\mathbf{Q}$ of the generalized problem introduced in Appendix A:

$$\hat{L} \cdot \mathbf{Q} = \omega \mathbf{M}_{\mathrm{g}} \cdot \mathbf{Q} + i\mathbf{j}_{\mathrm{g}} , \qquad \text{with} \quad \mathbf{j}_{\mathrm{g}} = \begin{pmatrix} \mathbf{j} & 0 & \ldots \end{pmatrix}^T . \tag{B7}$$

From the analysis of Appendix A, it is evident that $\mathbf{Q} = \begin{pmatrix} \mathbf{f} & \mathbf{Q}^{(1)} & \ldots \end{pmatrix}^T$. Furthermore, the eigenmodes of the generalized problem [Eq. (A9)] can be decomposed as $\mathbf{Q}_n = \begin{pmatrix} \mathbf{f}_n & \mathbf{Q}_n^{(1)} & \ldots \end{pmatrix}^T$ and $\mathbf{Q}_n^{\mathrm{c}} = \begin{pmatrix} \mathbf{f}_n^{\mathrm{c}} & \mathbf{Q}_n^{\mathrm{c},(1)} & \ldots \end{pmatrix}^T$. Thus, Eq. (B4) implies that

$$\mathbf{f}(\mathbf{r}) = \sum_n \mathbf{f}_n(\mathbf{r}) c_n , \qquad \text{with} \quad c_n = \left\langle \mathbf{Q}_n^{\mathrm{c}} \mid \mathbf{Q} \right\rangle . \tag{B8}$$

In Appendix C, we prove that $\langle \mathbf{Q}_n^{\mathrm{c}} | \mathbf{Q} \rangle$ can be expressed in terms of the electromagnetic fields as [Eq. (C5)]:

$$\langle \mathbf{Q}_n^{\mathrm{c}} | \mathbf{Q} \rangle = \frac{1}{2}\int_V d^3\mathbf{r}\, \mathbf{f}_n^{\mathrm{c}*}(\mathbf{r}) \cdot \left( \frac{\omega_{\mathrm{c},n}^* \mathbf{M}(\omega_{\mathrm{c},n}^*) - \omega \mathbf{M}(\omega)}{\omega_{\mathrm{c},n}^* - \omega} \right) \cdot \mathbf{f}(\mathbf{r})\ . \tag{B9}$$

Using Eq. (A1) and Eq. (B6), the above identity can be expressed as:

$$c_n = \frac{1}{\omega_n - \omega}\frac{1}{2}\int_V d^3\mathbf{r} \left[\hat{\mathcal{D}}\cdot \mathbf{f}_n^{\mathrm{c}}\right]^* \cdot \mathbf{f}(\mathbf{r}) - \mathbf{f}_n^{\mathrm{c}*}\cdot\left[\hat{\mathcal{D}}\cdot\mathbf{f} - i\mathbf{j}\right]. \tag{B10}$$

We used the equalities $\omega_{\mathrm{c},n}^* = \omega_n$ and $\mathbf{M}(\omega^*) = \left[\mathbf{M}^{\mathrm{c}}(\omega)\right]^\dagger$. As $\hat{\mathcal{D}}$ is Hermitian with respect to the canonical inner product, the above formula yields the desired result [Eq. (B5)].

The bi-orthogonal conditions in Eq. (B3) can be rewritten directly in terms of the electromagnetic modes using again Eq. (C5). Specifically, the electromagnetic modes must be normalized as:

$$\frac{1}{2}\int_V d^3\mathbf{r}\ \mathbf{f}_m^{\mathrm{c}*}(\mathbf{r}) \cdot \left( \frac{\omega_m \mathbf{M}(\omega_m) - \omega_n \mathbf{M}(\omega_n)}{\omega_m - \omega_n} \right) \cdot \mathbf{f}_n(\mathbf{r}) = 0, \qquad \text{if } \omega_n \neq \omega_m \tag{B11a}$$

$$\frac{1}{2}\int_V d^3\mathbf{r}\ \mathbf{f}_m^{\mathrm{c}*}(\mathbf{r}) \cdot \left( \frac{\partial}{\partial\omega}\left[\omega \mathbf{M}(\omega)\right] \right)_{\omega=\omega_n} \cdot \mathbf{f}_n(\mathbf{r}) = \delta_{n,m}, \qquad \text{if } \omega_n = \omega_m \tag{B11b}$$

These formulas generalize the results of Refs. [28, 29] to non-Hermitian systems.

***Modal Expansion of the Photonic Green's function***

Let $\overline{\mathcal{G}}(\mathbf{r},\mathbf{r}',\omega)$ (6×6 tensor) be the system Green's function, defined as in Eq. (8) of the main text. From Eq. (B5), it follows that the Green's function can be expanded in terms of the system normal modes as:

$$\overline{\mathcal{G}}(\mathbf{r},\mathbf{r}',\omega)=\frac{\omega}{2}\sum_n \frac{1}{\omega_n-\omega}\mathbf{f}_n(\mathbf{r})\otimes\mathbf{f}_n^{c*}(\mathbf{r}'), \tag{B12}$$

The electromagnetic modes satisfy the bi-orthogonal normalization conditions expressed in Eq. (B11).

## Appendix C: Inner Product Expressed in Terms of Electromagnetic Fields

In this Appendix, we prove the identity (B9). To this end, let us consider generic time-harmonic solutions $\mathbf{Q}(\mathbf{r},t)=\tilde{\mathbf{Q}}(\mathbf{r})e^{-i\omega t}$ and $\mathbf{Q}^{c}(\mathbf{r},t)=\tilde{\mathbf{Q}}^{c}(\mathbf{r})e^{-i\omega_c t}$ of the source-driven generalized problem [see Eqs. (A4) and (A6)] associated with the operator $\hat{L}$ and $\hat{L}^{c}$, respectively, so that

$$\hat{L}\cdot\tilde{\mathbf{Q}}=\omega\mathbf{M}_{g}\cdot\tilde{\mathbf{Q}}+i\tilde{\mathbf{j}}_{g}, \qquad \hat{L}^{c}\cdot\tilde{\mathbf{Q}}^{c}=\omega_{c}\mathbf{M}_{g}\cdot\tilde{\mathbf{Q}}^{c}+i\tilde{\mathbf{j}}_{g}^{c}. \tag{C1}$$

Here, $\tilde{\mathbf{Q}}$ and $\tilde{\mathbf{Q}}^{c}$ are the envelopes of the state vectors, and $\tilde{\mathbf{j}}_{g}=(\mathbf{j}\quad 0\quad ...)^{T}$ and $\tilde{\mathbf{j}}_{g}^{c}=(\mathbf{j}^{c}\quad 0\quad ...)^{T}$ are the envelopes of the generalized currents. The electromagnetic currents $\mathbf{j}$ and $\mathbf{j}^{c}$ are arbitrary and unrelated.

The state vector envelopes can be decomposed as $\tilde{\mathbf{Q}}=\left(\mathbf{f}\quad \mathbf{Q}^{(1)}\quad ...\right)^{T}$ and $\tilde{\mathbf{Q}}^{c}=\left(\mathbf{f}^{c}\quad \mathbf{Q}^{c,(1)}\quad ...\right)^{T}$. From the analysis of Appendix A, it should be evident that the corresponding electromagnetic field projections, $\mathbf{f}$ and $\mathbf{f}^{\mathrm{c}}$, are solutions of the source-driven frequency domain Maxwell's equations:

$$\hat{\mathcal{D}}\cdot\mathbf{f}=\omega\mathbf{M}(\mathbf{r},\omega)\cdot\mathbf{f}+i\mathbf{j}, \qquad \hat{\mathcal{D}}\cdot\mathbf{f}^{\mathrm{c}}=\omega_{\mathrm{c}}\mathbf{M}^{\mathrm{c}}(\mathbf{r},\omega_{\mathrm{c}})\cdot\mathbf{f}^{\mathrm{c}}+i\mathbf{j}^{\mathrm{c}}. \tag{C2}$$

Furthermore, from Eq. (A3), $\mathbf{Q}^{(\alpha)},\mathbf{Q}^{\mathrm{c},(\alpha)}$ can be expressed as:

$$\mathbf{Q}^{(\alpha)}=\frac{\left(s_\alpha\omega_{p,\alpha}\right)^{1/2}}{\left(\omega-\omega_{p,\alpha}\right)}\mathbf{A}_\alpha\cdot\mathbf{f}, \qquad \mathbf{Q}^{\mathrm{c},(\alpha)}=\frac{\left(s_\alpha\omega_{p,\alpha}^{*}\right)^{1/2}}{\left(\omega_{\mathrm{c}}-\omega_{p,\alpha}^{*}\right)}\mathbf{A}_\alpha^{\dagger}\cdot\mathbf{f}^{\mathrm{c}}. \tag{C3}$$

Consequently, the weighted inner product of the two state vector envelopes, $\left\langle\tilde{\mathbf{Q}}^{\mathrm{c}}\middle|\tilde{\mathbf{Q}}\right\rangle$, can be written only in terms of the corresponding electromagnetic field projections. Indeed, explicit calculations show that:

$$\begin{aligned}\tilde{\mathbf{Q}}^{\mathrm{c}*}\cdot\mathbf{M}_{\mathrm{g}}\cdot\tilde{\mathbf{Q}}&=\mathbf{f}^{\mathrm{c}*}\cdot\mathbf{M}_{\infty}\cdot\mathbf{f}+\sum_\alpha\mathbf{f}^{\mathrm{c}*}\cdot\frac{-\omega_{p,\alpha}\mathbf{R}_\alpha}{\left(\omega_{\mathrm{c}}^{*}-\omega_{p,\alpha}\right)\left(\omega-\omega_{p,\alpha}\right)}\cdot\mathbf{f}\\&=\mathbf{f}^{\mathrm{c}*}\cdot\left(\frac{\omega_{\mathrm{c}}^{*}\mathbf{M}\left(\omega_{\mathrm{c}}^{*}\right)-\omega\mathbf{M}(\omega)}{\omega_{\mathrm{c}}^{*}-\omega}\right)\cdot\mathbf{f}.\end{aligned} \tag{C4}$$

In the second identity we have used the partial-fraction expansion (A2). From the above formula, it follows that $\left\langle\tilde{\mathbf{Q}}^{\mathrm{c}}\middle|\tilde{\mathbf{Q}}\right\rangle$ can be written in terms of the electromagnetic components as in Eq. (B9),

$$\left\langle\tilde{\mathbf{Q}}^{c}\middle|\tilde{\mathbf{Q}}\right\rangle=\frac{1}{2}\int_V d^3\mathbf{r}\,\mathbf{f}^{\mathrm{c}*}(\mathbf{r})\cdot\left(\frac{\omega_{\mathrm{c}}^{*}\mathbf{M}\left(\omega_{\mathrm{c}}^{*}\right)-\omega\mathbf{M}(\omega)}{\omega_{\mathrm{c}}^{*}-\omega}\right)\cdot\mathbf{f}(\mathbf{r}), \tag{C5}$$

as we wanted to show. When $\omega=\omega_{\mathrm{c}}^{*}$, the term inside brackets should be replaced by $\frac{\partial}{\partial\omega}\left[\omega\mathbf{M}(\omega)\right]$.

## Appendix D: Proof of Eq. (10)

In this Appendix, we provide an explicit proof of Eq. (10) using the modal expansion in Eq. (14). Evidently, it is enough to demonstrate that $\nabla\cdot\mathbf{S}_n(\mathbf{r};\alpha)=0$ where $\mathbf{S}_n(\mathbf{r};\alpha)$ is defined as in Eq. (13b).

To this end, we note that:

$$
\begin{aligned}
& i\nabla\cdot\left\{\mathbf{E}_n(\mathbf{r};\alpha)\times\mathbf{H}_n^*(\mathbf{r};-\alpha)+\mathbf{E}_n^*(\mathbf{r};-\alpha)\times\mathbf{H}_n(\mathbf{r};\alpha)\right\} \\
& =-\mathbf{f}_n^*(\mathbf{r};-\alpha)\cdot\hat{\mathcal{D}}\cdot\mathbf{f}_n(\mathbf{r};\alpha)+\left[\hat{\mathcal{D}}\cdot\mathbf{f}_n(\mathbf{r};-\alpha)\right]^*\cdot\mathbf{f}_n(\mathbf{r};\alpha)
\end{aligned}
\tag{D1}
$$

where $\hat{\mathcal{D}}$ is the differential operator defined in Eq. (9). Next, we use [see Eq. (B6)]:

$$
\hat{\mathcal{D}}\cdot\mathbf{f}_n=\omega_n\mathbf{M}(\mathbf{r},\omega_n)\cdot\mathbf{f}_n, \qquad \hat{\mathcal{D}}\cdot\mathbf{f}_n^{\mathrm{c}}=\omega_{\mathrm{c},n}\mathbf{M}^{\mathrm{c}}\left(\mathbf{r},\omega_{\mathrm{c},n}\right)\cdot\mathbf{f}_n^{\mathrm{c}}. \tag{D2}
$$

where $\mathbf{f}_n$ ($\mathbf{f}_n^{\mathrm{c}}$) stands for $\mathbf{f}_n(\mathbf{r};\alpha)$ ($\mathbf{f}_n(\mathbf{r};-\alpha)$), respectively, and $\omega_{\mathrm{c},n}=\omega_n^*$. We took into account that $\mathbf{f}_n(\mathbf{r};-\alpha)$ is associated with the Hermitian conjugated problem, described by a material matrix such that $\mathbf{M}^{\mathrm{c}}(\omega)=\left[\mathbf{M}(\omega^*)\right]^{\dagger}$ [Eq. (A7)]. Substituting Eq. (D2) into Eq. (D1), one finds that:

$$
\nabla\cdot\mathbf{S}_n(\mathbf{r};\alpha)=\frac{1}{2i}\left[-\omega_n\mathbf{f}_n^{\mathrm{c}*}\cdot\mathbf{M}(\mathbf{r},\omega_n)\cdot\mathbf{f}_n+\omega_n\mathbf{f}_n^{\mathrm{c}*}\cdot\left[\mathbf{M}^{\mathrm{c}}\left(\mathbf{r},\omega_{\mathrm{c},n}\right)\right]^{\dagger}\cdot\mathbf{f}_n\right]=0 \tag{D3}
$$

In the rightmost identity we used $\left[\mathbf{M}^{\mathrm{c}}(\mathbf{r},\omega^*)\right]^{\dagger}=\mathbf{M}(\mathbf{r},\omega)$. Therefore, the divergence of the heat current in Eq. (14) vanishes, as we wanted show.

## Appendix E: Field singularities in a quasi-static approximation

In this Appendix, we characterize the field singularities near the wedge depicted in Fig. 3, using a quasi-static approximation. The wedge consists of two regions: an air sector and magnetized ferrite sector, surrounded by metallic plate.

We focus the analysis is characterization of waves of the type $\mathbf{E} = E_z(x,y)\hat{\mathbf{z}}$ and $\mathbf{H} = H_x(x,y)\hat{\mathbf{x}} + H_y(x,y)\hat{\mathbf{y}}$, which is consistent with the structure of the TE modes in a ferrite loaded rectangular waveguide (specifically the TE*m0* modes) [1]. Furthermore, for convenience we shall adopt a system of cylindrical coordinates $(\rho,\varphi,z)$ centred at the wedge vertex.

***Local model***

To begin with, we consider that the magnetized ferrite is modeled by the permeability tensor in Eq. (1), so that its response is local. We regard the parameters $\mu_t$, $\mu_g$ in Eq. (1) as position dependent, so that they can model both the magnetized ferrite and the air region. Evidently, in the air region we have $\mu_t = \mu_g = \varepsilon_r = 1$.

From the Maxwell's equations, the magnetic field in polar coordinates ($\mathbf{H} = H_\rho\hat{\boldsymbol{\rho}} + H_\varphi\hat{\boldsymbol{\varphi}}$) can be written in terms of the electric field as:

$$H_\rho = \frac{1}{i\omega\mu_0}\frac{1}{\mu_t^2-\mu_g^2}\left(\mu_t\frac{1}{\rho}\partial_\varphi E_z - i\mu_g\partial_\rho E_z\right), \tag{E1a}$$

$$H_\varphi = \frac{1}{i\omega\mu_0}\frac{1}{\mu_t^2-\mu_g^2}\left(-\mu_t\partial_\rho E_z - i\mu_g\frac{1}{\rho}\partial_\varphi E_z\right). \tag{E1b}$$

We used $\mathbf{H} = \frac{1}{i\omega}\overline{\mu}^{-1}\cdot\nabla\times\mathbf{E}$ with $\nabla\times\mathbf{E} = \nabla E_z\times\hat{\mathbf{z}} = \frac{1}{\rho}\partial_\varphi E_z\hat{\boldsymbol{\rho}} - \partial_\rho E_z\hat{\boldsymbol{\varphi}}$. Next, we impose that $\nabla\times\mathbf{H} = -i\omega\varepsilon\mathbf{E}$ in each homogeneous region, leading to:

$$\frac{1}{\rho}\partial_\rho\left[\rho\partial_\rho E_z\right] + \frac{1}{\rho^2}\partial_\varphi^2 E_z + \left(\frac{\omega}{c}\right)^2\varepsilon_r\mu_{\mathrm{ef}}E_z = 0, \tag{E2}$$

where $\mu_{\rm ef} = \dfrac{\mu_t^2 - \mu_g^2}{\mu_t}$. Note that the wave equation in each homogeneous region is insensitive to the gyrotropic response.

In order to characterize the field singularities near the wedge, we neglect time retardation, so that the speed of the light is taken as $c = \infty$. This leads to:

$$\frac{1}{\rho}\partial_\rho\left[\rho\partial_\rho E_z\right] + \frac{1}{\rho^2}\partial_\varphi^2 E_z \approx 0, \qquad \text{(E3)}$$

which is precisely the Laplace equation in cylindrical coordinates. We can find its solutions considering an ansatz of the type $E_z = \rho^\nu g(\varphi)$. This results in

$$\partial_\varphi^2 g + \nu^2 g = 0. \qquad \text{(E4)}$$

Thus, in each homogeneous angular region of the type $\varphi_i \le \varphi \le \varphi_{i+1}$ the azimuthal field profile satisfies:

$$g(\varphi) = A_i \cos\left(\nu\left(\varphi - \varphi_i\right)\right) + B_i \sin\left(\nu\left(\varphi - \varphi_i\right)\right), \qquad \text{(E5)}$$

for some constants $A_i, B_i$. The field components $E_z$ and $H_\rho$ are required to satisfy continuity boundary conditions across the material interfaces $\varphi = \varphi_i$:

$$g \text{ and } h \text{ are continuous.} \qquad \text{(E6)}$$

We introduced the function:

$$h \equiv \frac{1}{\mu_t^2 - \mu_g^2}\left(\mu_t \partial_\varphi g - i\mu_g \nu g\right). \qquad \text{(E7)}$$

After some calculations, it can be shown that the values of $g$ and $h$ calculated at the interfaces $\varphi = \varphi_{i+1}$ and $\varphi = \varphi_i$ are related as $\begin{pmatrix} g_{i+1} \\ h_{i+1} \end{pmatrix} = \mathbf{T}_i \begin{pmatrix} g_i \\ h_i \end{pmatrix}$, with $\mathbf{T}_i$ the transfer matrix:

$$\mathbf{T}_i = \begin{pmatrix} \cos\left(\nu\,\delta\varphi_i\right) + \dfrac{i\mu_{g,i}}{\mu_{t,i}}\sin\left(\nu\delta\varphi_i\right) & \dfrac{1}{\nu}\dfrac{\mu_{t,i}^2 - \mu_{g,i}^2}{\mu_{t,i}}\sin\left(\nu\delta\varphi_i\right) \\ -\nu\dfrac{1}{\mu_{t,i}}\sin\left(\nu\delta\varphi_i\right) & \cos\left(\nu\,\delta\varphi_i\right) - \dfrac{i\mu_{g,i}}{\mu_{t,i}}\sin\left(\nu\delta\varphi_i\right) \end{pmatrix}, \tag{E8}$$

with $\delta\varphi_i = \varphi_{i+1} - \varphi_i$. The index $i$ in the material parameters identifies the components of the permeability tensor within the homogeneous sector $\varphi_i \le \varphi \le \varphi_{i+1}$.

Next, we use the developed formalism to characterize the transfer matrix for the geometry shown in Fig. 3. Specifically, the global transfer matrix for this system is determined by a product of the transfer matrix associated with the air sector ($\mathbf{T}_{\text{air}}$) and the magnetized ferrite sector ($\mathbf{T}_{\text{F}}$): $\mathbf{T}_{\text{glob}} = \mathbf{T}_{\text{F}} \cdot \mathbf{T}_{\text{air}}$. Both transfer matrices are evaluated with $\delta\varphi_i = \pi/2$. In order that the electric field vanishes at $\varphi = 0$ and $\varphi = \pi$ (the metallic walls), it is necessary that $\left(\mathbf{T}_{\text{glob}}\right)_{12} = 0$. After some simplifications, it can be shown that this equation is satisfied when:

$$\cot\left(\frac{\pi}{2}\nu\right) = \frac{-i\mu_g}{\mu_t^2 - \mu_g^2 + \mu_t} \qquad \text{or} \qquad \sin\left(\frac{\pi}{2}\nu\right) = 0. \tag{E9}$$

The parameters on the right-hand side of the first equation are the permeability components of the magnetized ferrite.

The solutions of the second equation are evidently $\nu = 2n$ with $n$ a nonzero integer (the solution $n = 0$ is discarded because it is associated with a trivial magnetic field). As explained in the main text, these solutions are insensitive to the material response, and hence are of no interest to us.

The coefficient $\nu$ determines the singular behavior of the fields near the wedge vertex. Indeed, using $E_z = \rho^\nu g\left(\varphi\right)$ and Eq. (E1), one can readily show that:

$$E_z \sim \rho^{\nu}, \qquad \mathbf{H} \sim \rho^{\nu-1}. \tag{E10}$$

Next, discuss how $\nu = \nu(\omega;\alpha)$ varies with the damping factor $\alpha$. To this end, we start by noting that from Eq. (A7) the Hermitian conjugate permeability response is such that $\overline{\mu}(\omega;-\alpha) = \left[\overline{\mu}(\omega^*;\alpha)\right]^{\dagger}$. This identity can also be verified by direct substitution in Eq. (1). In particular, the permeability components obey $\mu_t(\omega;-\alpha) = \mu_t^*(\omega^*;\alpha)$ and $\mu_g(\omega;-\alpha) = \mu_g^*(\omega^*;\alpha)$.

On the other hand, the asymptotic behavior of the fields near wedge vertex is determined by the solutions of Eq. (E9). Conjugating both sides of the equation and using $\mu_t(\omega;-\alpha) = \mu_t^*(\omega^*;\alpha)$ and $\mu_g(\omega;-\alpha) = \mu_g^*(\omega^*;\alpha)$, it can be readily shown that the singularity exponent satisfies:

$$\nu(\omega^*;-\alpha) = -\nu^*(\omega;\alpha). \tag{E11}$$

Thus, we have demonstrated that the asymptotic behavior of the fields of the Hermitian conjugate problem (with gain, $\alpha < 0$) is fully determined by the asymptotic behavior of the fields in the original system (with dissipation, $\alpha > 0$).

***Nonlocal model***

Next, we study the asymptotic behavior of the magnetic field when the ferrite is described by the nonlocal model introduced in Sect. IVA [Eqs. (23), (25) and (26)].

From Eq. (23), the electric field satisfies a wave equation:

$$\nabla^2 E_z = -i\omega\mu_0 \hat{\mathbf{z}} \cdot \nabla \times (\mathbf{H} + \mathbf{M}). \tag{E12}$$

Time-harmonic variation of the fields is implicit. In the homogeneous ferrite sector, the magnetization vector satisfies:

$$\left[-\frac{1}{k_{\max}^2}\nabla^2+1\right]\mathbf{M}=\begin{pmatrix}\mu_t-1 & -i\mu_g & 0\\ i\mu_g & \mu_t-1 & 0\\ 0 & 0 & 1\end{pmatrix}\cdot\mathbf{H}, \tag{E13}$$

with all components of $\mathbf{M}$ subject to Dirichlet boundary conditions at the ferrite interfaces. Here, $\mu_t,\mu_g$ represent the permeability components of the local model. The above formula follows from Eqs. (25)-(26), noting that $\mathbf{M}$ and $\langle\mathbf{H}_{\mathrm{loc}}\rangle$ in the nonlocal approach are related in the same way as $\mathbf{M}$ and $\mathbf{H}$ in the local model.

From the theory of elliptic operators, $\mathbf{M}$ is smoother than $\mathbf{H}$ near the wedge vertex, because the solution of Eq. (E13) "gains" two derivatives in regularity relative to the independent term on the right-hand side. Consequently, $\mathbf{M}$ is always more regular than $\mathbf{H}$ in the ferrite sector, and thereby it can be dropped in the wave equation [Eq. (E12)].

The wave equation reduces then to $\nabla^2 E_z \approx -i\omega\mu_0\hat{\mathbf{z}}\cdot\nabla\times\mathbf{H}=-\left(\frac{\omega}{c}\right)^2\varepsilon_r E_z\approx 0$, where we neglected retardation effects ($c\to\infty$) and used Eq. (23). Thus, analogous to the local problem, we find that $\nabla^2 E_z\approx 0$, i.e. the field behavior near the wedge vertex is still governed by the Laplace equation. Furthermore, the magnetic field behavior near the wedge vertex is still determined by $\mathbf{H}\sim\nabla\times\mathbf{E}$ because as noted before the magnetization vector is smoother than $\mathbf{H}$.

The key difference between the local and nonlocal formalisms lies in the boundary conditions for $\mathbf{H}$ at the ferrite–air interface. In the local case, the continuity of the tangential magnetic field at the air-ferrite boundary implies the continuity of $\hat{\boldsymbol{\rho}}\cdot\overline{\mu}^{-1}\cdot\nabla\times\mathbf{E}$, which is equivalent to Eq. (E7), as discussed earlier.

In the nonlocal case, the relevant boundary condition can be found noting that $\mathbf{H} = -\mathbf{M} + \frac{1}{i\omega\mu_0}\nabla\times\mathbf{E}$ [Eq. (23)]. Since $\mathbf{M}$ vanishes at the ferrite boundary, the continuity of the magnetic field at the ferrite-air boundary reduces to the continuity of $\hat{\boldsymbol{\rho}}\cdot\nabla\times\mathbf{E}$. Thus, unlike in the local case, the introduction of nonlocality renders the boundary conditions insensitive to the material response of the ferrite. As a result, the singularity exponents $\nu$ coincide with those of the reciprocal case, i.e., $\nu$ is an integer.

## Appendix F: Eigenmodes of the reciprocal dual system

In this Appendix, we establish the relationship between the eigenmodes of the reciprocal dual system and those of the Hermitian-conjugate problem. As in the main text, we denote the family of eigenmodes of the original system by $\mathbf{f}_n(\mathbf{r};\alpha,\mathbf{B}_0)$ and the corresponding eigenfrequencies by $\omega_n(\alpha,\mathbf{B}_0)$. Since the eigenmodes are solutions of Maxwell's equations without sources, they satisfy [Eq. (B6)]

$$\hat{\mathcal{D}}\cdot\mathbf{f}_n(\mathbf{r};\alpha,\mathbf{B}_0) = \omega_n(\alpha,\mathbf{B}_0)\mathbf{M}(\mathbf{r},\omega_n;\alpha,\mathbf{B}_0)\cdot\mathbf{f}_n(\mathbf{r};\alpha,\mathbf{B}_0). \tag{F1}$$

Using $\left[\boldsymbol{\sigma}\cdot\hat{\mathcal{D}}\cdot\boldsymbol{\sigma}\right]^* = \hat{\mathcal{D}}$ and $\boldsymbol{\sigma}^2 = \mathbf{1}_{6\times6}$, it follows from Eq. (F1) that:

$$\hat{\mathcal{D}}\cdot\boldsymbol{\sigma}\cdot\mathbf{f}_n^*(\mathbf{r};\alpha,\mathbf{B}_0) = \omega_n^*(\alpha,\mathbf{B}_0)\tilde{\mathbf{M}}\cdot\boldsymbol{\sigma}\cdot\mathbf{f}_n^*(\mathbf{r};\alpha,\mathbf{B}_0). \tag{F2}$$

with $\tilde{\mathbf{M}} = \boldsymbol{\sigma}\cdot\mathbf{M}^*(\mathbf{r},\omega_n;\alpha,\mathbf{B}_0)\cdot\boldsymbol{\sigma}$.

From Eq. (A7), the Hermitian conjugated system has a material response such that $\mathbf{M}^{c}(\omega) = \left[\mathbf{M}(\omega^*)\right]^\dagger$. Therefore, we have $\mathbf{M}^*(\mathbf{r},\omega_n;\alpha,\mathbf{B}_0) = \mathbf{M}^T(\mathbf{r},\omega_n^*;-\alpha,\mathbf{B}_0)$, which implies that $\tilde{\mathbf{M}} = \boldsymbol{\sigma}\cdot\mathbf{M}^T(\mathbf{r},\omega_n^*;-\alpha,\mathbf{B}_0)\cdot\boldsymbol{\sigma}$. But from Eq. (19), $\boldsymbol{\sigma}\cdot\mathbf{M}^T(\mathbf{r},\omega_n^*;-\alpha,\mathbf{B}_0)\cdot\boldsymbol{\sigma}$ corresponds to the material matrix of a reciprocal dual system, i.e., a system with a flipped magnetic bias. Thus, the preceding analysis shows that $\tilde{\mathbf{M}} = \mathbf{M}\left(\mathbf{r}, \omega_n^*; -\alpha, -\mathbf{B}_0\right)$. Substituting this result into Eq. (F1), we see that:

$$\hat{\mathcal{D}} \cdot \boldsymbol{\sigma} \cdot \mathbf{f}_n^*\left(\mathbf{r}; \alpha, \mathbf{B}_0\right) = \omega_n^*\left(\alpha, \mathbf{B}_0\right) \mathbf{M}\left(\mathbf{r}, \omega_n^*; -\alpha, -\mathbf{B}_0\right) \cdot \boldsymbol{\sigma} \cdot \mathbf{f}_n^*\left(\mathbf{r}; \alpha, \mathbf{B}_0\right). \quad \text{(F3)}$$

This identity demonstrates that $\boldsymbol{\sigma} \cdot \mathbf{f}_n^*\left(\mathbf{r}; \alpha, \mathbf{B}_0\right)$ are the eigenmodes of the reciprocal dual of the Hermitian conjugate problem, i.e., the system with flipped magnetic bias and negative dissipation. Specifically, we have:

$$\mathbf{f}_n\left(\mathbf{r}; -\alpha, -\mathbf{B}_0\right) = \boldsymbol{\sigma} \cdot \mathbf{f}_n^*\left(\mathbf{r}; \alpha, \mathbf{B}_0\right), \qquad \omega_n\left(-\alpha, -\mathbf{B}_0\right) = \omega_n^*\left(\alpha, \mathbf{B}_0\right). \quad \text{(F4)}$$

As $\omega_n\left(\alpha, \mathbf{B}_0\right) = \omega_n^*\left(-\alpha, \mathbf{B}_0\right)$ [Eq. (B1)], the above result is equivalent to Eq. (20) of the main text.